\documentclass[USenglish,twocolumn]{article}

\usepackage[utf8]{inputenc}
\usepackage[big]{dgruyter_author}

\usepackage[dvipsnames,table]{xcolor}
\usepackage{flushend}
\usepackage{amsmath}
\usepackage{float}
\usepackage{amsthm}
\usepackage{placeins}
\usepackage{algorithmic}
\usepackage{array}
\usepackage{stfloats}
\usepackage{multirow}
\usepackage{multicol}
\usepackage{tabularx}
\usepackage{slashbox,pict2e}
\usepackage[switch]{lineno}
\usepackage{hyperref}
\usepackage{subcaption}
\definecolor{myblue}{RGB}{0, 126, 240}
\definecolor{ltblue}{RGB}{180,177,250}
\usepackage[math]{cellspace}
\makeatletter
\renewcommand\@makefntext[1]%
    {\noindent\makebox[0pt][r]{\textsuperscript{\@thefnmark}\,}#1}
\makeatother

\begin{document}

  \articletype{Research Article{\hfill}Open Access}

  \author*[1]{Yosef S. Razin}

\author[2]{Karen M. Feigh}

  \affil[1]{School of Aerospace Engineering,
Georgia Institute of Technology,
Atlanta, Georgia 30332--0250, E-mail: yrazin@gatech.edu}

  \affil[2]{School of Aerospace Engineering,
Georgia Institute of Technology,
Atlanta, Georgia 30332--0250, E-mail: karen.feigh@gatech.edu}

  \title{\huge Committing to Interdependence: Implications from Game Theory for Human-Robot Trust}

  \runningtitle{An Interdependence Model of Human-Robot Trust}


  \begin{abstract}
{Human-robot interaction and game theory have developed distinct theories of trust for over three decades in relative isolation from one another.  Human-robot interaction has focused on the underlying dimensions, layers, correlates, and antecedents of trust models, while game theory has concentrated on the psychology and strategies behind singular trust decisions. Both fields have grappled to understand over-trust and trust calibration, as well as how to measure trust expectations, risk, and vulnerability. This paper presents initial steps in closing the gap between these fields. Using insights and experimental findings from interdependence theory and social psychology, this work starts by analyzing a large game theory competition data set to demonstrate that the strongest predictors for a wide variety of human-human trust interactions are the interdependence-derived variables for commitment and trust that we have developed.  It then presents a second study with human subject results for more realistic trust scenarios, involving both human-human and human-machine trust.  In both the competition data and our experimental data, we demonstrate that the interdependence metrics better capture social `overtrust' than either rational or normative psychological reasoning, as proposed by game theory.  This work further explores how interdependence theory--with its focus on commitment, coercion, and cooperation--addresses many of the proposed underlying constructs and antecedents within human-robot trust, shedding new light on key similarities and differences that arise when robots replace humans in trust interactions.}
\end{abstract}

  \journalname{Paladyn, J. Behav. Robot.}

  \startpage{1}

\maketitle

\section{Introduction}
Human-robot interaction (HRI) and game theory have had little interaction in the development of their respective theories of trust and collaboration. Game theory has long utilized a singular concept of trust, defined as the payoff structure of typically one-shot interactions. It thereby attempted to figure out not what trust looked like behaviorally, but what psychological motivations led to its fulfillment \cite{Bacharach2007,dunning2014trust}.  Conversely, HRI focused primarily on deconstructing the idea of trust, its underlying dimensions, antecedents, and corollaries \cite{schaefer2013perception,Madsen2000}. This attempt to understand trust more holistically, as a system of attitudes, expectations, decisions, and behaviors, led to many insights at the cost of construct proliferation. Beyond this conceptual rift between the disciplines, HRI often viewed game theory's `games' as derived from toy problems that didn't translate well into the field; this was despite HRI's own trust research being often limited to simulations or 2D interfaces with accompanying post-task questionnaires$^{\tiny \footnotemark}$\footnotetext{For more on this divide and its general implications see \cite{Carp2004}}. These divisions and approaches can be traced back to the origins of these parallel paths of exploring trust.  

Trust in HRI has been strongly influenced by social psychology, human factors, and teamwork, whereas trust in game theory has been more strongly influenced by philosophy, economics, and political science.  While both fields have drawn liberally from others and have independently developed their own unique insights, they have yet to cross-germinate fruitfully. This paper will begin to bridge that gap, starting a new conversation on what HRI (and human-machine interaction more generally) can learn from the primarily human-human interactions studied in game theory and looking at how the underlying constructs of trust from HRI relate to game-theoretic trust.

In this paper, we will first present an overview of trust as it's been approached by HRI and game theory, as well as the current tenuous connections between the two fields. We will then give a brief introduction to interdependence theory, focusing on how it contributes to our understanding of trust by deriving a testable definition of trust games, and developing equations for commitment and a new trust index. We then present two experiments: the first testing interdependence theory-based algorithms on a game theory competition data set followed by our own human subject testing, showing the power of interdependence theory over previously proposed approaches to trust prediction. Finally, we will discuss the power as well as the limits of our approach, especially with regard to human-human vs. human-machine trust.

\subsection{Trust and Control in HRI}
\label{sec:types}
Early work focusing on trust in automation mainly grew out of social psychology \cite{Deutsch1958,deutsch1960effect,deutsch1977resolution,Rotter_67,Rotter1980,Rempel1985}. It was also influenced by sociology, primarily Niklas Luhmann’s Trust and Power \cite{luhmann} and Bernard Barber’s The Logic and Limits of Trust \cite{barber1983logic} proved to be hugely influential, firmly establishing trust as multi-dimensional, and explicating its relation with complexity and communication. Luhmann's influence can still be identified in two major disputes within HRI trust, as far as the roles of norms and control \cite{jalava2003}. Briefly, does the modern world and its complex technologies, such as robotics, with their inherent uncertainties and risks, preclude familiarity and norm-based trust? Furthermore, are systems of control replacements for trust in such a world instead of an integral part of trust itself?  While Luhmann answered both of these in the affirmative, these questions are currently coming to the fore of debates in HRI trust. How we answer these questions will have profound implications, especially for how HRI trust is conceived in contrast to human-human trust.

While generic trust had been historically captured by a single item on survey instruments, once trust was understood as multidimensional and distinct from confidence and familiarity, early researchers of trust in automation started trying to capture these new dimensions \cite{Muir1987,Lee_94}. Eventually, some of these axes converged around a slightly shortened form of Mayer’s seminal definition of trust, as
\begin{quote} “the willingness of a party to be vulnerable to the actions of another party based on the expectation that the other will perform a particular action important to the trustor” \end{quote} that is rooted in the constructs of ‘Ability’, ‘Integrity’, and ‘Benevolence’  \cite{Mayer1995}. Later works created more fully fleshed out trust models, antecedents, and co-factors, incorporating concepts such as interface design, understandability, transparency, ease of use, effectiveness, accessibility, and familiarity \cite{Madsen2000,Gefen2003,Korber2004,McKnightD2011,Chien2014,Hoff2015}.  Further developments included expanding Mayer's `Benevolence' into more general expectations concerning affective trust (e.g. cooperation vs. competitiveness)\cite{castelfranchi,schaefer2013perception,Chien2014} and re-casting Mayer's `Integrity' into structural assurance \cite{Gefen2003,McKnightD2011}. Finally, in order to discern whether expectations of trustworthiness truly transformed into trust, considerations of intended and actual use were considered (based on \cite{Gefen2003}).
    
Mayer's original definition had included the final clause, “irrespective of the ability to monitor or control that party”, in which Luhmann's dichotomy of external control in opposition to trust can be discerned.  This clause has often been dropped in later HRI trust definitions (\textit{e.g.} \cite{Lee2004}), however more recent works have expanded and explicated what such control means.  Castelfranchi and Falcone \cite{castelfranchi} have argued that while narrow, `strict' trust is antagonistic to control, a broader notion of trust that includes confidence in social systems and norms, such as laws, contracts, and ethics actually completes and compliments trust, increasing it above what strict trust alone would suggest.  Similarly, Law and Scheutz \cite{law2020} understand trust as two distinct categories, performance-based and relation-based. Performance-based trust is relying on competence \textit{sans} monitoring (`strict'), whereas relation-based trust expands beyond the specific situation. This latter category hews closely to Luhmann's confidence in social systems of trust as well as Castelfranchi and Falcone's concept of `broader' trust.  A similar treatment of this `new', abstracted, social/normative `category' of trust is termed `structural trust' and found to be a well-defined, independent, and internally consistent dimension of HMI trust by McKnight \cite{McKnightD2011}, Gefen \cite{Gefen2003}, and Malle \& Ulman \cite{malle2020}. 

Thus, while the topology of trust is still contended, a consensus has emerged regarding the role of broader control via trust in social structures such as ethics and laws in HRI. This may be somewhat surprising, applying expectations of norms to robots. However, norms are a crucial part of familiarity and expectation building \cite{luhmann}, even for non-human agents. One clear example is autonomous vehicles.  For instance, Razin \& Feigh \cite{razin2020} demonstrated that while drivers rated human and autonomous vehicles differently based on perceived performance-based trust, they had the same expectations and perceptions of both agents when it came to social expectations around driving. In other words, they believed that self-driving cars would follow the same laws and norms as humans on the road.  Similar results on the importance of norms to trust around household interactions have also been identified \cite{salem2015towards,salem2015would}. The human trustor may also abstractly place structural trust in a company that sells them robotic products, the engineers that design those products, and the laws that regulate the products and businesses involved \cite{Gefen2003,McKnightD2011}.  Structural trust devolves upon an entire social network of actors of which the actual robot is only one node, albeit the facet at which the direct trust interaction occurs.

Beyond control, cooperation in the form of teamwork is receiving increased attention in HRI \cite{schaefer2013perception,atkinson2014shared} as is coercion (both through incentives and sanctions), especially in the form of inappropriate compliance and reliance \cite{meyer2014measures,parasuraman2010complacency}.  While this work focuses on these larger questions of control, cooperation, coercion, and commitment, we will return to discussing how performance-based and affective-based trust fit into the interdependence model (See Sec. \ref{sec:HRI_imp}).  In the following, we will also differentiate `rational' trust from performance-based `strict' trust given the specific meaning of rationality in game theory; indeed, the bulk of this work is aimed at explaining that both performance-based and affective-based trust are based on rational beliefs.

\subsection{Interdependence Theory: Deconstructing Control}
\begin{figure}
    \centering
    \includegraphics[width=\columnwidth]{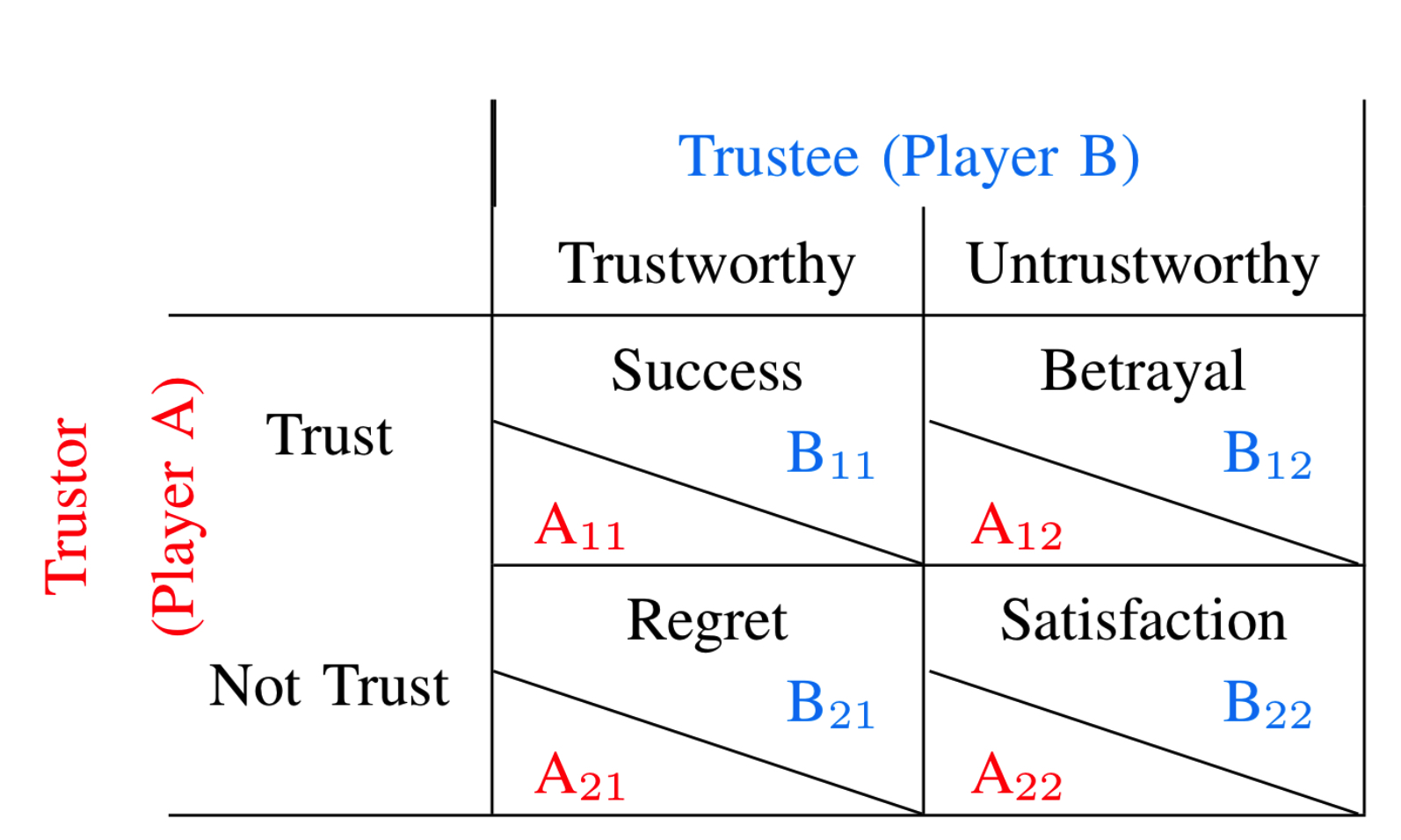}
    \caption{Payoff matrix for the trustor (red) and trustee (blue) in a trust-trustworthiness interaction. Regret here is specific to not trusting/being trusted when trust would have been fulfilled and is distinct from any emotion linked to betrayal.}
   \label{fig:outcomeMatrix}
\end{figure}

In order to further explore the relationship of trust with control, cooperation, and coercion, we propose reviving an off-shoot of game theory, proposed by Kelley \& Thibaut over a half-century ago \cite{thibaut_59}. Their interdependence theory was reintroduced into HRI trust by Wagner \cite{Wagner2009} and Robinette \cite{robinette} and re-frames classical games by breaking down the relative levels of control afforded to each agent.  The theory of interdependence is also broader than classical game theory as it does not assume rationality or even the attempt to maximize monetary or even concrete outcomes. Thus, it considers symbolic outcomes, such as the reputational payoff of following social norms or the pleasure of fulfilling another's needs \cite{VanLange}.  Kelley \& Thibaut also recognized that even within a single interaction, the `game' is not limited to simply the structure of the outcomes prescribed by the situation, but that actors may further process and mentally transform such situations by framing them in various temporal or social ways.  These include attempting to maximize the joint outcomes of all actors or minimizing the difference between some outcomes to ensure equity.  They also explored transformations instantiated by making certain externally-motivated behavioral commitments (playing by the rules, turn-taking), preempting partner's choices, and accounting for future interactions \cite{thibaut_59}.  These transformations act in well-characterized and prescribed ways upon outcome matrices similar to those used in game theory, such as the 2x2 matrix in Fig. \ref{fig:outcomeMatrix}. Many of these steps to expand game theory would be retread starting in the late 1980s within mainstream game theory research by Geanakoplos, Pearce, \& Stacchetti \cite{GPS}, when they founded psychological game theory. However, the focus on the decomposition of games by control `modes' remains a unique and crucial contribution of interdependence theory alone.

\subsection{Trust in Game Theory}
Before turning to the method of deconstructing control within a game or interaction, it is worthwhile understanding how trust is even framed in game theory, which is so often focused on competitive scenarios.  Unlike trust in HRI, in game theory trust is not seen as multidimensional and there is little debate over its definition. What defines a trust game in game theory, first and foremost, is the payoff structure (see Fig. \ref{fig:outcomeMatrix}). The oft-cited requirements (e.g. \cite{Ermisch2006,Bacharach2007,Balliet2013})
for a trust game according to game theory are

\begin{enumerate}
    \item Exposure: The trustor is risking more by betrayal than if they don't trust ($A_{12}<\{A_{21},A_{22}\}$)
    \item Improvement: The trustor stands to gain more by fulfilled trust than by not trusting ($A_{11}>\{A_{21},A_{22}\}$)
    \item Temptation: The trustee at least is tempted to betray trust when proffered ($B_{12}>B_{11}$)
    \item Mutual Gain$^{\tiny \footnotemark}$\footnotetext{Not universally accepted}: That the payoff for being trustworthy when trusted is higher than not being trustworthy at all ($B_{11}>\{B_{21},B_{22}\}$)
\end{enumerate}
 
all while assuming that $A_{21}=A_{22}$ and $B_{21}=B_{22}$.

A very similar, though expanded, set of trust conditions for the \textit{trustor} alone was independently derived by Wagner \cite{Wagner2009} as:
 
\begin{enumerate}    
\item The act of trust must occur in the face of uncertainty; the trustee cannot act before the trustor.
\item Only if the trustor chooses to trust does the trustee's action matter, such that the payoff for successful trust is higher than the potential loss if the trustee is untrustworthy. Quantitatively,  this means the difference between the payoffs for successful vs unsuccessful trust must be at some minimum ($\epsilon_1$) dependent, reflecting some risk ($A_{11}-A_{12}>\epsilon_1$\label{eq:1}) (Exposure)
\item The trustor's payoffs for not trusting are independent of the trustee,  such that the amount unrisked by not trusting is bounded by $\epsilon_2$, such that $|A_{21}-A_{22}|<\epsilon_2$\label{eq:2}.
\item Successful trust is the highest outcome and betrayal the lowest, with the non-trusting options bound by these two levels, such that $A_{11}>\{A_{21},A_{22}\}>A_{12}$\label{eq:3} (Improvement)
\item The trustor must believe that the probability of the trustee acting trustworthy is greater than some trust threshold ($p^A(\text{TW})>C$\label{eq:4}).
 \end{enumerate}

Note that the inequalities presented by game theory only define a trust game and not how the binary decision to trust or act trustworthy is made. Whereas Wagner attempts to provide, at least abstractly, such a criterion in (5). One such solution for calculating that decision threshold could be the game's mixed Nash equilibrium.  However, game theory generally has suggested that the `rational' solution here is the subgame perfect equilibrium (SPE) - which unfortunately and unrealistically predicts that trust should rarely occur, as being untrustworthy is the trustee’s weakly dominant \textit{rational} strategy. This is clearly not how trust plays out in the real world, where trust is frequently given and fulfilled. Thus, behavioral and psychological insights are sought to fill this gap.


It is a well-known phenomenon that, even in the `toy problems' presented in game theory experiments, people choose to trust and be trustworthy more than they seemingly `should' based on payoffs and risk aversion alone \cite{dunning2014trust}. Theories as to why range the gamut from long-term reputation keeping, conformity to moral norms, expecting and reciprocating kindness, guilt, and inequality aversion to name but a few,  with varying supporting findings in the game-theoretic trust literature \cite{Ert2011,Balliet2013,dunning2014trust,Ermisch2006,Battigalli2005}. Note that these theories \textit{all} fall under the `broader' notions of structural or relation-based trust, as discussed above in Sec. \ref{sec:types}.

The gap between game theory's and HRI's approaches to trust and how they are articulated, framed, motivated, and modeled is wide indeed.  In fact, the only common foundation to both approaches is that trust occurs when one is made vulnerable by exposure to risk and that it is premised on a  “particular action of importance to the trustor” \cite{Mayer1995}.   Game theory focuses on the binary trust decision, and, more often than not, HRI focuses on the continuously valued belief in or expectation of trust and trustworthiness. Furthermore, the very design of the game-theoretic implementation removes questions of capability, much less understandability and familiarity, and completely violates the evaluation of trust under situational normality, which are all stressed in HRI.  On the other hand, one could argue that by removing these correlates, game theory examines a `purer’ form of trust that goes beyond instrumentality \cite{dunning2014trust}.  This `pure’ trust though is also strongly understood to be rooted in exclusive elements of human-human interaction, focusing on equity, kindness, and moral normativity. This approach completely disregards how trust operates when a non-human is involved (beyond anthropomorphism) or even when trust is strongly premised on performance as opposed to relational concerns. Furthermore, HRI trust rarely concerns itself with the trustee's alternative payoffs, with the exception of \cite{Wagner2009}, directly challenging the Temptation criteria of trust games as formally defined by game theory.

\subsection{A Declaration of Interdependence}
\begin{figure*}
    \centering
    \begin{subfigure}[t]{0.45\textwidth}
    \includegraphics[width=0.95\columnwidth, trim = {0 0 0 0},clip]{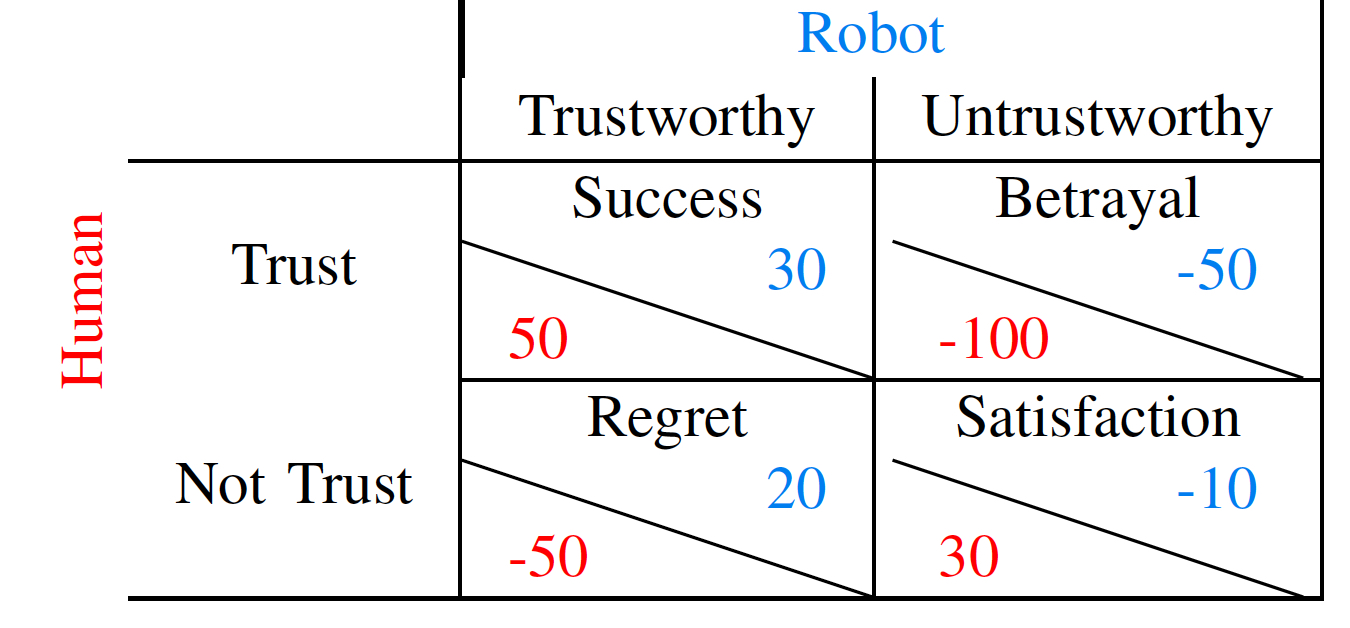}
    \caption{Example normal-form representation of the trust game, best suited for enabling calculations}
    \end{subfigure}
    \begin{subfigure}[t]{0.45\textwidth}
    \includegraphics[width=0.95\columnwidth, trim = {0 0 0 0},clip]{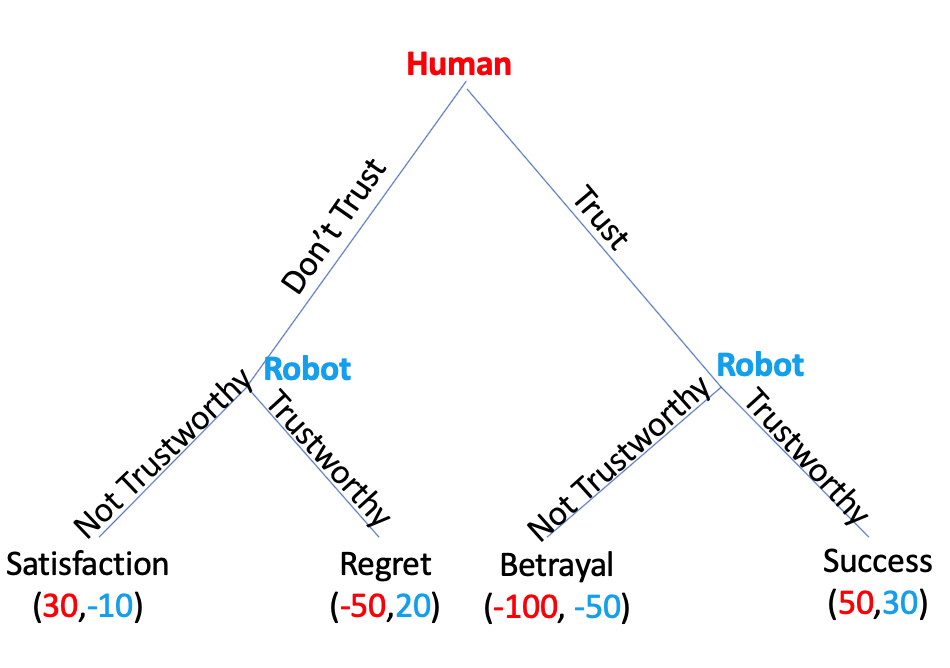}
    \caption{Extensive-form representation of the game, highlighting the sequential nature of the game, with the trustor making the first `move'}
    \end{subfigure}
    \caption{An illustrative example of the payoffs in a human-robot trust interaction. The human trustor's payoffs are in red and the robot's are in blue, with the game represented in two forms, (a) the normal-form payoff matrix and (b) the extensive-form game.}
    \label{fig:ext_game1}
\end{figure*}

Interdependence theory proposes decomposing games by determining which actor has power over which part of the total payoff structure \cite{thibaut_59}. We will illustrate each aspect of this powerful approach and its insights through the following game, shown in Fig. \ref{fig:ext_game1}.

Imagine a human must decide whether to trust an autonomous vehicle or switch to manual mode. The payoff structure here is not just the real costs or payoffs but also incorporates emotional, reputational, and other psychological utilities. If the human does trust and the autonomous vehicle works perfectly, the human is reasonably happy especially given the cost of the car (\textcolor{red}{50}).  If they don't trust the autonomous mode, even though they believe it generally works, and get into an accident by driving manually, they will kick themselves for not trusting and regret it (\textcolor{red}{-50}). However, if the human trusts the autonomous vehicle and it fails, it is catastrophic and they may never use the car again (\textcolor{red}{-100}). Finally, if they decide not to trust it and then hear that it actually does not work they will feel satisfied with their justified choice (\textcolor{red}{30}). Note that due to psychological factors such as regret and satisfaction, typically $A_{21}\neq A_{22}$.

The payoffs for the robot can be seen as the utility either for it as an agent directly or for its owners, manufacturers, designers, or insurers.  The robot anticipates being rewarded and used more (or perhaps its manufacturer anticipates increased share prices) for properly fulfilling trust (\textcolor{blue}{30}) but penalized even more if it betrays the human's trust, as failure may result in discontinued use, not to speak of reputational and commercial loss (\textcolor{blue}{-50}).  While often such games assume that the trustee receives or losses nothing by being not trusted ($B_{21}=B_{22}=0$) \cite{dunning2014trust,Wagner2009}, that is clearly not the case, as can be illustrated. Generally, not being trusted will hurt the brand (\textcolor{blue}{-10}) but as people get into accidents driving manually while the robot actually is demonstrably a more reliable driver than humans, then the safer autonomous vehicle will appear a better option and people will seek it out (let's say a net gain of \textcolor{blue}{20}). 

Traditional game theory would predict that the two agents will act rationally and play the subgame perfect equilibrium (SPE).  What this means is that by working backward through the example in Fig. \ref{fig:ext_game1}, the autonomous car's payoff for regret dominates satisfaction ($\textcolor{blue}{20}>\textcolor{blue}{-10}$) and successfully fulfilling trust dominates betrayal ($\textcolor{blue}{30}>\textcolor{blue}{-50}$). The human trustor is then left deciding between successful trust vs regret ($\textcolor{red}{50}>\textcolor{red}{-50}$), and thus in this case the SPE predicts that the human will indeed successfully trust.  However, in cases when satisfaction dominates regret, such as when the trustee is seen as less as a tool and more as a potential teammate, the SPE indicates that one should not trust. In practice, the SPE seems to account for approximately 60-80\% of trustor's decision to trust in human-human interaction \cite{Bacharach2007,Ert2011,dunning2014trust}.

Note that payoffs in game theory are known to be invariant under positive affine transformation - thus it does not matter if we multiply all the payoffs of the human by 1000 or add 50 to each of those of robot. It also makes it tricky (if not impossible) to compare payoffs between the agents. However, normalizing all payoffs by each player's most extreme outcome can prove useful for understanding interdependence, as will be shown shortly.


Interdependence theory suggests that we can understand the interaction better by deconstructing the payoffs in terms of three types of control, those of each individual as well as that which arises from cooperation.  Reflexive or actor control (RC) is how much unilateral power the actor has over their own outcomes, \textit{i.e.} the expected difference between their choosing one action over the other.  For the trustor (Player A), it is the average difference between the row-sums and for the trustee (Player B) this is transposed as the average difference between column-sums, such that
\vspace{-1mm}
\begin{align}
    RC_A = 0.5\big((A_{11}+A_{12})-(A_{21}+A_{22})\big)\;\\\nonumber
    RC_B = 0.5\big((B_{11}+B_{21})-(B_{12}+B_{22})\big).
\end{align}

In the human-robot game illustrated in Fig. \ref{fig:ext_game1}, the normalized reflexive control is RC$_A$= \textcolor{red}{-.15} for the human trustor (Player A) and RC$_B$= \textcolor{blue}{0.7} for the car trustee (Player B). Thus, along this component of the payoff, the human is weakly inclined to choose not to trust and the autonomous vehicle (and its manufacturer) has a much stronger incentive to prove trustworthy.

Fate or partner control (FC) is how much unilateral power each actor has over the other's outcomes, \textit{i.e.} the expected difference in one actor's outcomes when the other chooses between their actions.  For Actor A, it is the average difference between their payoff's column-sums, and again this is transposed for Actor B:
\vspace{-0.5mm}
\begin{align}
    FC_A = 0.5\big((A_{11}+A_{21})-(A_{12}+A_{22})\big)\;\\\nonumber
    FC_B = 0.5\big((B_{11}+B_{12})-(B_{21}+B_{22})\big).
\end{align}
To reiterate, FC$_A$ is Actor A's estimation of Actor B's unilateral power over A's own outcomes. Per our example, FC$_A$=\textcolor{red}{0.35} and FC$_B$=\textcolor{blue}{0.1} meaning the vehicle's trustworthiness has a much stronger impact on the human, than the human's choice to trust the car (RC$_A$= \textcolor{red}{-.15}), while the driver/consumer provides a mild positive car/manufacturer for the vehicle to be trustworthy.

Finally, bilateral or joint control (BC) is how much one actor's choice further facilitates or inhibits the other's outcomes.  This set of weights is fully contingent on the partner's choice and, thus, is the result of coordination or its lack thereof.  It is calculated for both partners as the average of the difference of the sums of the diagonal outcomes.
\vspace{-0.5mm}
\begin{align}
BC_X=0.5\big((X_{11}+X_{22})-(X_{12}+X_{21})\big),
\end{align}
where $X$ can be either $A$ or $B$.  Again in our interaction example, BC$_A$ = \textcolor{red}{1.15} and BC$_B$ = \textcolor{blue}{0.9}. Both the human and the car have a strong incentive to coordinate, fulfilling trust when trusted or not trusting the untrustworthy. Here control over payoffs via coordination is significantly stronger than any via unilateral control.

\begin{table}[h!]
    \centering
    \begin{tabular}{l|c|c}
        \textbf{Control Mode} & Human (A) & Car (B)  \\\hline
        Reflexive Control (RC) &\textcolor{red}{-.15}  & \textcolor{blue}{0.7}\\
        Fate Control (FC) &\textcolor{red}{0.35}  & \textcolor{blue}{0.1} \\
        Bilateral Control (BC) &\textcolor{red}{1.15}  & \textcolor{blue}{0.9}\\
    \end{tabular}
    \caption{Summary of Interdependence control modes from the human-robot interaction example based on normalized payoffs.}
    \label{tab:my_label}
\end{table}

If the signs of BC$_A$ and BC$_B$ are the same, as in the example, they are said to \textit{correspond}, signaling that both actors share a preference for coordinated behavior.  If the sign of BC matches that of RC or FC, they are said to be \textit{concordant}, and if not, they are \textit{discordant}.  \textit{Concordance} (\textit{discordance}) is a measure of reinforcement (interference) between one mode of control and another. As the autonomous vehicle's interdependence weights in our example are all positive, $FC_B$, $RC_B$, and $BC_B$ are all concordant, whereas for the human $FC_A$ and $BC_A$ are concordant but $RC_A$ is discordant with both, signifying that the human is being coerced (in this case through incentivization). $BC_A$ and $BC_B$ are both positive and thus correspond, indicating that the human and car share a preference for coordination.

In this way, interdependence theory and its associated weights can be used for a variety of analyses that illuminate many aspects of trust. This is clearly illustrated when we translate the game-theoretic trust conditions of Exposure and Improvement (or equivalently Wagner's trust conditions) into interdependence terms$^{\tiny \footnotemark}$\footnotetext{For proofs, see Appendix A.2: Theorems 1-4} as 
\vspace{-.5mm}
\begin{align}
\begin{split}
     FC_A &> 0 \hspace{1cm} FC_A > |RC_A| \\
     BC_A &> 0 \hspace{1cm} BC_A > |RC_A|
     \end{split}
    \label{eq:8a}
\end{align}

This transformation yields the following interpretation 
\begin{quote}
the expected additional gains of trusting that the trustee controls for the trustor, both unilaterally and through coordination, must be positive and greater in magnitude than the trustor's power over their own payoffs. 
\end{quote}
Note how this accords with our commonsense understanding of trust: if the trustor can accomplish the goal for themselves, there is no need for trust. Furthermore, the trustee exerts control over the trustor's success - and this arises from a mixture of coordination and magnanimity. This does not, however, imply altruism, as these conditions say nothing concerning the payoffs for the trustee. If we accept game theory's conditions for the trustee, Temptation can be expressed as:
\begin{align}
\begin{split}
     FC_B &> BC_B\\
     FC_B &> RC_B
     \end{split}
    \label{eq:8b}
\end{align} 

If the Mutual Gain condition is also accepted,
\begin{align}
\begin{split}
     FC_B &> |BC_B|\\
     FC_B &> |RC_B|
     \end{split}
    \label{eq:8c}
\end{align} 

Recall that often in game theory, trust games are designed such that $B_{21}=B_{22}$ \& $A_{21}=A_{22}$, which leads to 
\begin{align}
    FC_A=BC_A\\
    FC_B=BC_B,
\end{align}

however, neither of these two equivalence conditions are held to be actual requirements of trust games.

Note that the Temptation condition implies a somewhat cynical approach. Under it, a trust game is not simply when the trustor might consider trust as a viable strategy but when they would do so at the same time as when the trustee is tempted to betray them!  From interdependence theory, we see that this essentially means that the only time they will act trustworthy in such a scenario is when the trustor's unilateral control provides an overwhelming incentive.  While this may make for a `good' game, is it true that trust can only be said to occur in the presence of temptation/when it is coerced?

Leaving that question to be addressed below, observe how the interdependence analysis has allowed us to go beyond the basics of the game-theoretic conditions, highlighting the cooperative aspect, as well as the power inequality between the two agents \cite{razin2019}.  

There's another gain to note from this new framing of the trust conditions. As mentioned above, payoffs in game theory cannot be compared between agents because they are invariant under positive affine transformations. These trust conditions contain the additional benefit of permitting normalized ranges for the interdependence weights, such that RC$_A$=(-1,1), FC$_A$=(0,1), BC$_A$=(0,2), RC$_B$=(-2,1), FC$_B$=(-2,2), and BC$_B$=(-2,1). Thus, various trustors and their valuations can be fruitfully compared, and likewise for trustees. 

Finally, we recognize that in this section we have introduced a number of key terms and acronyms and we will be introducing more in the following sections. Thus, for ease of reference, a glossary of terms is provided in Appendix A.1.

\section{From Interdependence Weights to Measures to Trust}
Armed with these re-framed theoretical constraints, it is time to forge a new path to show how trust is actually decided upon within such interactions.  The initial starting points offered by game theory would be the subgame perfect equilibrium (SPE) and mixed Nash equilibrium. As previously mentioned, the SPE has proven insufficient at capturing actual trust behavior for the trustor, leading to what appears to be over-trust. One solution may be the Nash Equilibrium which broadens our view from a discrete decision to the continuous domain of probabilities.

The Nash equilibrium is the point of indifference between the trustee's actions given the trustor's payoffs and vice versa.  The probability of the trustee acting trustworthy ($\tau_B$), which yields a Nash equilibrium if the trustee is to use a mixed strategy, is derived$^{\tiny \footnotemark}$\footnotetext{For proof, see Appendix, Theorem B.1} as\\
\begin{align}\label{eq:10}
\begin{split}
    &\tau_BA_{11}+(1-\tau_B)A_{12} =   \tau_BA_{21}+(1-\tau_B)A_{22}\\
     &\Rightarrow \tau_B = \frac{A_{22}-A_{12}}{A_{11}+A_{22}-A_{12}-A_{21}}= \frac{1}{2}-\frac{RC_A}{2BC_A}.\\
     \\
    \end{split}
\end{align}

If the trustor is willing to assume that the trustee is rational, they can use $\tau_B$ as a best-response threshold to make decisions based on the trustee's trustworthiness ($p^A(TW)>\tau_B$; where $\tau_B=C$ in Wagner's 5\textsuperscript{th} condition).  While the basic result is well-known \cite{nash}, we can still glean a few key insights. First, the Nash equilibrium only holds if the trustee is assumed to be `rational'.  Of course, if one suspects the other as being `nasty' or a direct competitor, then the assumption of rationality does not hold \cite{gottman2011science}. Secondly, this approach highlights that a single act of untrustworthiness or trustworthiness may not be meaningful, but that trustworthiness is to be assessed dynamically over the relationship's span or at least from some previous expectation or likelihood, shedding light on the roles of familiarity and learning in trust.  However, it does not address `thin', one-shot trust interactions, though a Bayesian \textit{prior} over the trustor's belief may be considered as a potential alternative.

\subsection{In Gottman's Index, Trust}
Unhappy with the poor accuracy of the Nash equilibrium to predict trust and its limitation to rational actors, Gottman proposed a trust index based on his findings from experimental psychology \cite{gottman2011science}. This index was based on an idea from the Nash equilibrium that we want to maximize the payoffs such that no player can unilaterally choose a move that does better, but drops the rationality assumption and is neither predicated on interaction history nor the probability of trust.  While Gottman's original trust index was based on three potential actions per actor, we present a modified version of it here limited to the binary trust decision and translated into interdependence terms$^{\tiny \footnotemark}$\footnotetext{For proofs, see Appendix, Theorem A.3}. The trust index, TI, is thus given by:
\vspace{-0.5mm}
\begin{align}\label{eq:11}
\begin{split}
   TI &= \frac{A_{11}-A_{22}}{A_{11}+A_{21}-A_{12}-A_{22}}= \frac{1}{2}+\frac{RC_A}{2FC_A},\\
    \end{split}
\end{align}
where we recall that RC$_A$ is the unilateral control the trustor has over their own outcomes and FC$_A$ is the unilateral control the trustee has over the trustor's outcomes. Gottman describes this as ``without regard for the trustee's gains, the trustee can be counted on to look out for the trustor's interests by changing their behavior to improve the trustor's outcomes" \cite{gottman2011science}.  We can also understand this index as the equilibrium achieved between the trustor's choices given the probability, TI, that the actors will not match behaviors: trust will meet untrustworthiness and distrust with trustworthiness, as it can be derived from
\vspace{-0.5mm}
\begin{equation}
    (1-TI)A_{11}+TI A_{12} = (1-TI)A_{22}+TIA_{21}.
\end{equation}

Given our derived constraint $FC_A>|RC_A|$ in Eq. \ref{eq:3}, the index is no longer arbitrary, as presented by Gottman, but becomes a proper metric, such that a trust interaction can be said to not exist if TI$<0$ or TI$>1$. Furthermore, when $0.5<$TI$<1$ trust can be said to be freely given, and $0<$TI$<0.5$ trust is forced or coerced.  The latter could occur if RC$_A$ and FC$_A$ are \textit{discordant} and since FC$_A$ >0 (see Eq. \ref{eq:3}), trust must be being incentivized by the trustee against the trustor's negative inclination (RC$_A<0$, FC$_A>0$) \cite{razin2019}, as in the example above with the autonomous car where TI$=0.29$. One question explored further below, is whether this incentivization/coercion is enough to convince the trustor to go ahead despite their misgivings.  Gottman validated his trust index through its positive correlation with the trustor's higher emotional attunement and lower physiological arousal and the trustee's reduced negativity and greater openness during oral relationship history interviews. Thus, he concluded, his trust index does indeed reflect trust within intimate relationships \cite{gottman2011science}. As we will show, the index will also prove to be a powerful tool for predicting trust in both human-human and human-machine interactions.





\subsection{Committing Trust}
Is the trust index alone sufficient to predict trust? It seems to capture much of the interplay in the one-on-one interaction. However, it does not seem to address a central question of trust researchers from both HRI \cite{Lee_94} and psychology \cite{rusbult_93}, on how one decides to interact in the first place.  Often we think of trust as a choice between doing something ourselves versus delegating to another; this has been studied in HRI classically as the self-confidence vs human-machine trust going back to Lee \& Morray \cite{Lee_94}. However, as others have pointed out \cite{castelfranchi}, we often have more than one potential trustee - whether we are choosing between apps, a new car, lab partners, potential business opportunities, or people to date.  How do we choose which of these avenues are worth pursuing?

In their initial work on interdependence theory, Kelly and Thibaut introduced the idea of the comparison level for the alternative, CL$_{alt}$ \cite{thibaut_59}.  This is a set point from which we compare our lowest acceptable payoffs for each interaction. The lower CL$_{alt}$, the more the interaction is worth pursuing among the set of all interactions; CL$_{alt}$ connotes the anticipated worth of the current interaction and the likelihood that it will be pursued further (\textit{i.e.} the interaction's stability).

Caryl Rusbult developed this idea further with Kelley \cite{rusbult_93}, successfully validating the idea that CL$_{alt}$ can be understood psychologically as one's commitment to the interaction.  She found in her Investment Model of Commitment that as people discount or reject other CL$_{alt}$'s in favor of the current relationship, they become more invested in and dependent on the relationship. Likewise, increasing the worth of CL$_{alt}$ by comparing other potential partners to one's current partner, for instance, leads to a cascade not just of distrust but ultimately betrayal. Gottman continued to build on and test this idea \cite{gottman2011science}, concluding that conflict avoidance exacerbates CL$_{alt}$, which is reflected in further detachment.  He also differentiated between commitment and trust, whereby ``turning away erodes trust" but ``turning away and increasing CL$_{alt}$ erodes trust and fuels betrayal" \cite{gottman2011science}. 

\begin{figure}
    \centering
    \includegraphics[width=\columnwidth,trim={0 0 0 15pt},clip]{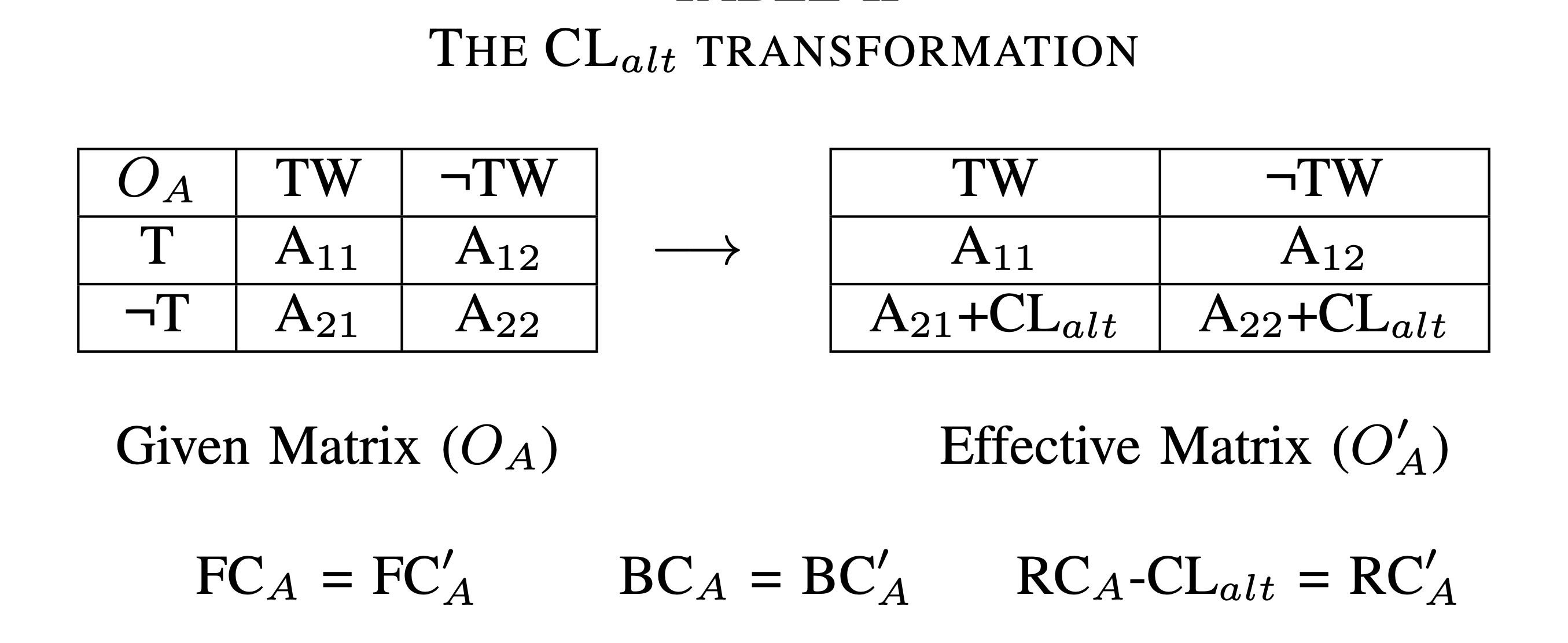}
    \caption{The CL$_{alt}$ transformation. CL$_{alt}$ only reduces RC$_A$ and therefore commitment}
    \label{fig:clalt}
\end{figure}

Based on the functional requirements for CL$_{alt}$ as described in the works above, we proposed a new transformation process \cite{razin2019}, like those in \cite{kelley_78}, to apply CL$_{alt}$ to an interaction and understand it via interdependence, as shown in Fig. \ref{fig:clalt}. The CL$_{alt}$ transformation does not affect FC$_A$ or BC$_A$ and thus does not directly affect the trustor's interdependence. However, increasing CL$_{alt}$ decreases RC$_A$ by an equivalent amount (\textit{i.e.} RC$_A'$ = RC$_A$-CL$_{alt}$). This transformation explains the erosion of Gottman's trust index as the trustor's commitment lessens and the ratio of RC$_A'$:FC$_A'$ decreases. 

As the trustor's commitment wanes, the percentage of the time the trustee must act trustworthy to `prove themselves' increases, as can be shown from Eq. \ref{eq:10}.  The idea of `neediness' in psychological game theory \cite{Bacharach2007} is mathematically equivalent to increasing RC$_A$ and commitment through consideration of a negative CL$_{alt}$ but is only mentioned in passing and is less developed therein.

As explained initially by Kelley \& Thibaut, the higher one actor's CL$_{alt}$ relative to the other's, the more power they are said to have in the relationship, though this is not necessarily true for any single interaction.  This is because one may choose to make themselves vulnerable (or needy as seen through the lens of game theory) in the short term, either through sacrifice or accommodation, in order to signal trustworthiness, without compromising their overall power. In the long term, however, doing so abdicates power and deepens one's dependence, commitment, and, indeed, `neediness'.   
In considering alternatives, there are two further effects that we have previously derived\cite{razin2019}, that are worth summarizing here. Given that the payoff/cost of the alternative, CL$_{alt}$, is inversely proportional to BC$_A$ in the Nash equilibrium and FC$_A$ in the Trust Index.

\begin{enumerate}
    \item As the cost of alternatives grows very high ($\Rightarrow$ CL$_{alt}<-2$FC$_A$), the commitment, RC'$_A$, increases to the point such that the trust index, $\Rightarrow$TI > 1 increases above and beyond what Player B's trustworthiness indicates. This can lead to a coerced over-trust by Player A through `sunk cost' or over-commitment.
    \item Strong alternatives (CL$_{alt}>>0$) decrease commitment, RC'$_A$, to the interaction at hand, lowering the expectation of trustworthiness, $\tau_B$, such that it may no longer meet the required threshold $C$ to trust, where $\Delta\tau_B=\frac{\text{CL}_{\text{alt}}}{\text{2BC}_A}$.
\end{enumerate}

Note that the first point provides a psychological, interdependence-based explanation of the sunk cost `fallacy'.  Here though it is perfectly rational and not a fallacy, per se.  Commitment is a sunk cost as other opportunities are foregone and more personal power is ceded so that the other must be increasingly trusted.

The last point is related to a concept that Gottman entitles `turning toward/away'.  Recall that BC is the payoff for cooperation.  It turns out that, at least theoretically, the higher the payoff for `turning toward’ the other, the lower the effect of alternatives should be. In other words, `turning away’ from the other decreases the robustness of relationships to alternatives, and `turning toward' the other increases its robustness, precisely the effect found in Gottman’s studies \cite{gottman2011science}.


\section{Experiment 1: Capturing Human-Human Trust}
While less fully developed, other works had previously noted the importance of coordination, commitment, and the trust index in theory and experimentation, both from human-human interaction \cite{gottman2011science} as well as, to a more limited extent, from human-robot trust \cite{Wagner2009}.  However, these concepts still lacked direct validation based upon quantitative data. Therefore, that is the first experimental goal of this work.  The second goal is to look at the implications of our findings and indicate further directions that such game-theoretic analysis may apply to HRI and in what ways it is expected to differ from human-human trust.

\subsection{Experimental Procedure 1}
To test and validate our work, we turned to a competition data set \cite{Ert2011} that contained 240 unique, non-trivial games generated from 10 ``classical" non-trivial game types, such as `trust', `near dictator', `costly punishment', and `safe shot'.  Each of the 240 generated games was played between 116 students that were paired off, but blind to each other’s choices, with pairings changed for each game. Students were drawn from a business school subject pool and compensated based on the payoffs and choices made in one of the played games, chosen at random. The games were divided into 120 for training the estimation algorithms and 120 to be used to validate prediction accuracy. Results from the top 15 performing algorithms in both the estimation and prediction components of the competition were publicly reported as well as baseline results and a coding template for implementation \cite{Ert2011}. 

To this data set, which included over a dozen strategies of gameplay, we added our various interdependence theory-derived variables as well as the trust index previously mentioned. The full list of  algorithms and variables can be found in Table \ref{tab:algs} and Table \ref{tab:vars}, respectively. We normalized all payoffs by their most extreme value, as discussed above, to counter issues that could arise given game theory's utility invariance under strictly positive affine transformations. After validating the baseline code, all games that did not fit our minimal criteria for defining trust games (e.g. exposure and improvement, Eq. \ref{eq:3}) were removed, resulting in a reduced set of 47 estimation games and 59 prediction games.  The data sets were not re-equalized by size, so that comparisons could be made against the baseline results from the competition. Games that did not fulfill the Temptation criteria were retained, in part due to previous HRI work not including that requirement in trust interactions \cite{Wagner2009,razin2019} and furthermore because temptation is directly related to the trustee's commitment, a condition which we wished to test and not simply exclude.  All parameter values in the baseline algorithms were reoptimized with the goal of minimizing the mean squared error. Our models did not make the strong presumption of \cite{Ert2011} to remove the intercept \textit{a priori}, since there was no reason to believe that the mean of trust on the y-axis should be 0. In fact, if trust is examined independently of any antecedents such as familiarity and faith in society, then it must account somewhere for potential background bias, which is at least expected on the part of the trustor. All regression algorithms were 10-fold cross-validated. 

\begin{table}[t]
    \centering
    \begin{tabular}{SlSc}
        Algorithm & Strategy   \\\hline
        \parbox{3cm}{Subgame Perfect \\ Equilibrium (SPE)}
         & \parbox{4.5cm}{Players follow\\ `rational' strategies}    \\\hline
        Inequality Aversion \cite{Ert2011}  & \parbox{4.5cm}{
        Players avoid inequality but\\weight disadvantageous and \\advantegous inequality differently} \\\hline
        \parbox{3cm}{Equality Reciprocity\\ (ERC)\cite{Bolton1995}} & \parbox{4.5cm}{Mixing SPE,  gains from co-\\ordination (trustor), and tit-for-tat\\ (trustee). All material payoffs\\ being equal, players prefer\\ equal distribution}  \\\hline
        Charness-Rabin \cite{Rabin2020,Ert2011} &  \parbox{4.5cm}{Combining SPE with the idea \\  of fairness/kindness (tit-for-tat)} \\\hline
         ``Seven Strategies" \cite{Ert2011} &
         \parbox{4.5cm}{Regression analysis of\\ strategies that one or\\ both players may employ.\\ See Table \ref{tab:vars} for full list.}\\
             \end{tabular}
                     \caption{Previously proposed strategies of trusting and trust fulfillment}
                         \label{tab:algs}
\end{table}

An important caveat of this data set (and in fact all game-theoretic and interdependence-based games in the literature) is that $A_{21}=A_{22}$ and $B_{21}=B_{22}$, which implies, for those still following, that $FC_A=BC_A$ and $RC_B=BC_B$. Thus, the trustee's commitment is equal to their additional incentive to cooperate, and the control the trustee has over the trustee's payoffs is an even mixture of unilateral and joint control.

\subsection{Results}

Due to concerns of multicollinearity among the 16 variables the variance inflation factors (VIF) for the data set were checked (see Table \ref{tab:vars}). Gottman’s trust index (TI) and the commitment of the trustor (RC$_A$) showed a correlation of 97\% and the trustee’s subgame perfect equilibrium (b1) and the trustee's strategy of maximizing `niceness' (mn1) were heavily correlated at 93\%. The trustor’s subgame perfect equilibrium (ri) had medium strength correlations with both the trustor assuming a malicious trustee (maxmin) and the trustee’s commitment (RC$_B$) (51\% and 55\%, respectively). Given the importance of subgame perfect equilibria and our hypothesis, we dropped maxmin and tested both dropping RC$_A$ and TI, settling on TI as it showed stronger results, which brought all VIF below 3 except for ri (VIF$=3.65$).

    \begin{table}
    \centering
    \begin{tabular}{SlScScSc}
         Variable & Meaning & VIF init. & VIF final \\\hline
         &\parbox{2.5cm}{\centering ``Seven Strategy"\\ Variables \cite{Ert2011}}&&\\\hline
         ri& \parbox{2.5cm}{\centering Subgame perfect equilibrium\\ for trustor}&\textbf{4.48} & \textbf{3.65} \\\hline
         lev1& \parbox{2.5cm}{\centering Trustor maximizing self-payoffs\\ given total uncertainty} & \textbf{4.07}& 2.78\\\hline
         mm1& \parbox{2.5cm}{\centering Trustor maximizing\\ payoffs of weakest\\ player (kindness)}& 2.04& 1.84\\\hline
         maxmin&\parbox{2.5cm}{\centering Trustor maximizing payoff\\ assuming Player 2 is malicious} & \textbf{3.60}& \textcolor{Maroon}{Dropped}\\\hline
         jm1& Maximizing joint payoffs &2.97 & 2.10\\\hline
         ia1& \parbox{2.5cm}{\centering Minimizing payoff differences\\ (equality)}& 1.97& 1.54\\\hline
         b1& \parbox{2.5cm}{\centering Subgame perfect equilibrium\\ for trustee}&\textbf{10.58} & 1.71\\\hline
         mn1& \parbox{2.7cm}{\centering Trustee maximizing\\ trustor’s payoff\\ if rational choice\\ is indifferent}
         & \textbf{11.07}&\textcolor{Maroon}{Dropped}\\\hline
         mm2& \parbox{2.5cm}{\centering Trustee maximizing\\ payoffs of weakest\\ player (kindness)}&2.20 & 1.88\\\hline
         ia2&\parbox{2.5cm}{\centering Minimizing payoff differences\\ (equality)}& 2.02& 1.57\\\hline
         \\[1ex]&\textbf{Interdependence Variables}&&\\[1ex]\hline
         RC$_A$& Trustor's commitment & \textbf{25.75} & \textcolor{Maroon}{Dropped} \\\hline
         FC$_A$/BC$_A$& \parbox{3cm}{\centering Trustee’s unilateral\\ and joint control\\ over trustor} &2.47 &2.46 \\\hline
         RC$_B$/BC$_B$&\parbox{3cm}{\centering Trustee’s commitment\\ and joint control} & 1.68& 1.36\\\hline
         FC$_B$&\parbox{3cm}{\centering Trustor’s control\\ over trustee} & 2.21& 1.95\\\hline
         TI& Gottman's Trust Index & \textbf{27.59}&2.26 \\\hline
    \end{tabular}
    \caption{Seven strategies and interdependence variables and their initial and final variance inflation factors (VIF), after strong multicollinear variables were dropped}
    \label{tab:vars}
\end{table}

After our data was checked for various statistical assumptions, all of the game play strategies of \cite{Ert2011} and several machine learning regressions led to the results shown in Table \ref{tab:main_results}. As in \cite{Ert2011}, playing the subgame perfect equilibrium alone still accounted for 75.3\% of the variance in the trustor's trusting response and 97.6\% of the variance for the trustee's fulfillment of trust in the reduced data set.

Interestingly even with reoptimizing parameters, many of the methods tested by \cite{Ert2011} did not perform as well as or only slightly better than the subgame perfect equilibria for prediction when only trust games were analyzed. This was despite their sometimes significant improvement over the estimation set. Additionally, when the full set of seven strategies and interdependence variables were fitted and tested in various regression schemes, the seven strategies variables were almost always discarded by the models as insignificant with regard to the trustor. All of the best performing algorithms (linear regression, support vector machine (SVM), Gaussian process regression (GPR), and ensemble regression) showed that the interdependence terms better captured the likelihood of trust, both in terms of lowest error rates and fewest terms. While SVM and GPR prevent us from examining which variables were most impactful, we can use linear regression and the tree-based methods (regression tree and the boosted ensemble) to draw some meaningful conclusions.

\begin{table}
    \centering
    \begin{tabular}{Sl|c|c|c|c|l}
     &  \multicolumn{2}{c}{Estimation} & \multicolumn{2}{c}{Prediction}\\
      Method  & Trustor & Trustee & Trustor & Trustee\\\hline
      \parbox{2.5cm}{Subgame Perfect\\ Equilibrium (SPE)}  & 0.1288 &0.0184 & 0.0432 & 0.0065\\\hline
      Inequality Aversion & 0.0336 & 0.0249 & 0.0229 & 0.0071\\\hline
      \parbox{2.5cm}{Equality Reciprocity\\ (ERC)} & 0.0378 & 0.0176 & 0.0509 & \cellcolor{ltblue}\textbf{0.0057}\\\hline
      Charness-Rabin & 0.0729 & 0.0036 & 0.0626 & 0.0263\\\hline
      Seven Strategies & 0.0802 & 0.0035 & 0.0373 & 0.0077 \\\hline 
      Linear Reg. & 0.0183 & 0.0098 & \cellcolor{ltblue} \textbf{0.0218} & 0.0069\\\hline
      Reg. SVM  & \cellcolor{ltblue}\textbf{0.0075} & \cellcolor{ltblue}\textbf{0.0020} & 0.0263 & \cellcolor{ltblue}\textbf{0.0051}\\\hline
      Reg. Tree & 0.0144 & 0.0035 & \cellcolor{ltblue}\textbf{0.0210} & 0.0124\\\hline
      Gaussian Proc. Reg.
      & \cellcolor{ltblue}\textbf{0.0065} & \cellcolor{ltblue}\textbf{0.0021} & 0.0219 & \cellcolor{ltblue}\textbf{0.0057}\\\hline 
    Ensemble Reg.   & \cellcolor{ltblue}\textbf{0.0052} & \cellcolor{ltblue}\textbf{0.0021} & \cellcolor{ltblue}\textbf{0.0141} & 0.0077\\\hline
    \end{tabular}
       \caption{Mean squared error for trust and trust fulfillment: All regressions were run at least initially with all Seven Strategy and interdependence variables. The three best performers along each category are highlighted.}
    \label{tab:main_results}
\end{table}

\begin{figure}[h]
    \centering
    \includegraphics[width=0.98\columnwidth, trim = {2pt 0 55pt 5pt},clip]{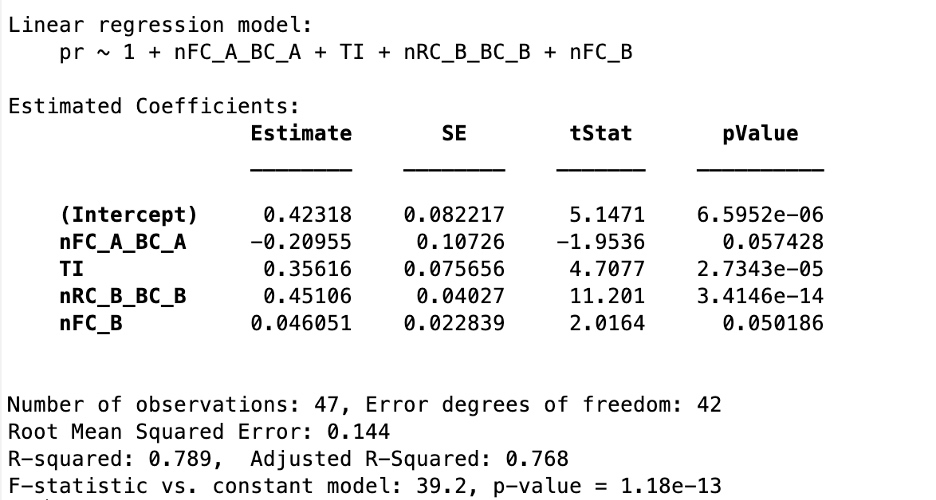}
    \caption{Final 10-fold cross-validated and stepwise-improved linear regression for the trustor. nFC\_A\_BC\_A is FC$_A$/BC$_A$, nRC\_B\_BC\_B is RC$_A$/BC$_B$, nFC\_B is FC$_B$, and $pr$ is the probability that trust was given.}
    \label{fig:lin1}
\end{figure}

\subsection{When to Trust}
Starting with the dropping of terms, as recommended by VIF, and then performing stepwise improvement, the final linear regression model (shown in Fig. \ref{fig:lin1}) found that there was a significant bias toward trusting (0.423) and that the most important variables were Gottman’s Trust Index (TI) and the trustee’s commitment/cooperation (RC$_B$/BC$_B$).  Both FC terms were of borderline significance (p=0.057 and 0.050).  Marginal improvement in the mean squared error occurred if FC$_A$/BC$_A$ was dropped (0.0244 to 0.0218), but there were no gains if FC$_B$ was removed.  Since all variables are normalized, the regression weights can be compared against each other-- indicating that while FC$_B$ may have borderline significance, its effect is an order of magnitude weaker than the other terms.

\begin{figure}[h]
    \centering
    \includegraphics[width=\columnwidth, trim = 5cm 3cm 6cm 5cm,clip]{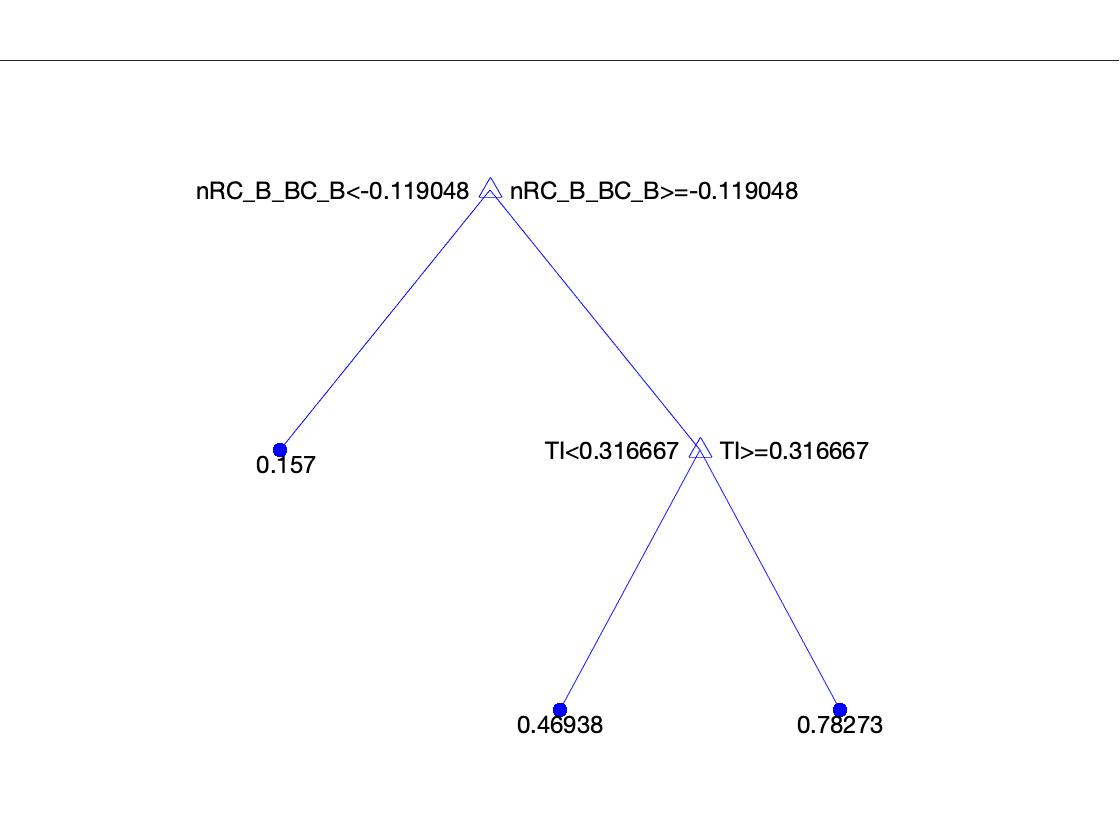}
    \caption{Boosted ensemble tree for trustor - the two retained variables are the trustee's commitment/joint control, RC$_B$/BC$_B$, and the trust index, TI.}
    \label{fig:ens1}
\end{figure}

The best performer, the least-squares boosted ensemble, showed similar results to the linear regression analysis, with the trustee’s commitment/cooperation gains (RC$_B$/BC$_B$) playing the largest role followed by the Trust Index (TI), as shown in Fig. \ref{fig:ens1}. Note that as long as the trustee seems at least indifferent to commitment/cooperation (RC$_B$/BC$_B>-0.12$), TI is sufficient for indicating whether trust is bestowed.  Of further interest is that trust is bestowed even if the trust index is below 0.5. In that regime, RC$_A$ and FC$_A$ have opposite signs, indicating that trust is being forced; in the case of this data set, FC$_A$ is always positive, which means that when TI$<0.5$, RC$_A$ must be negative and trust is being incentivized. Since neither term can be greater than one, we also see that the trustor's negative commitment is no more than 0.37 (RC$_A>-0.37$), so the trustor's lack of commitment in such cases may be understood as bordering on indifference. To summarize, the trustee’s control over the trustor’s outcome greatly overrides lack of commitment as long as the incentive to trust/cooperate is about 2.7 times greater. Furthermore, the trustor is likely to strongly trust the trustee when their own commitment aligns with that of the trustee's incentivization/cooperation. Thus, for the trustee, not only does incorporating the interdependence results better predict the probability of trusting but it appears that just a small subset of the interdependence variables alone gives a more accurate, simpler, and common-sense model of trust, than the `seven strategies' or models of trust based on (in)equality or fairness.

\begin{figure}[h]
    \centering
    \includegraphics[width=\columnwidth, trim = {2pt 0 60pt 0},clip]{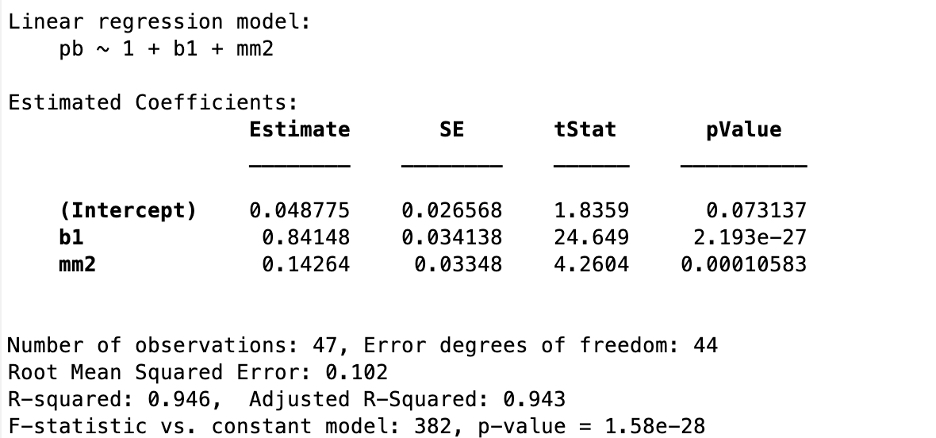}
     \caption{Final 10-fold cross-validated and stepwise-improved linear regression for the trustee. b1 is the trustee's subgame perfect equilibrium and mm2 is the trustee's strategy to help the weakest player.}
    \label{fig:lin2}
\end{figure}
\subsection{On Being Trusted}
\begin{figure}
    \centering
    \includegraphics[width=0.95\columnwidth, trim = 62pt 35pt 85pt 102pt,clip]{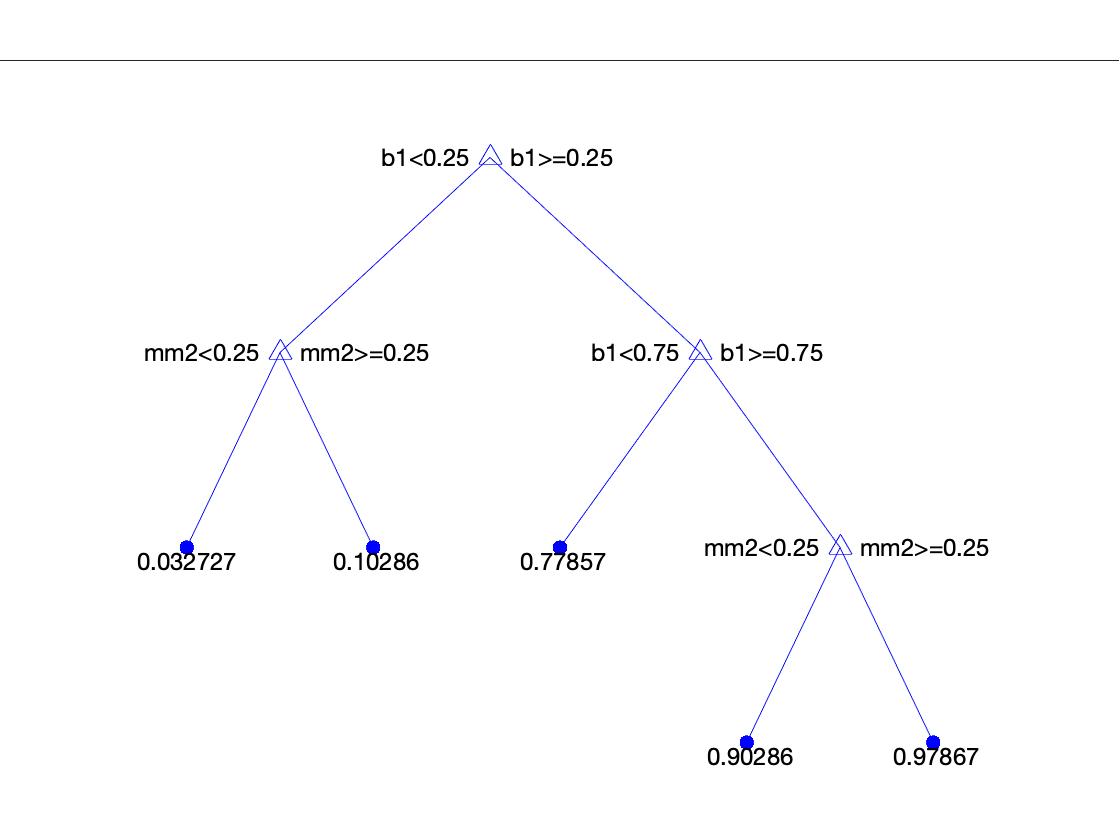}
    \caption{Regression tree for the trustee on deciding when to fulfill trust. b1 is the trustee's subgame perfect equilibrium and mm2 is the trustee's choosing to help the weakest player (kindness)}
    \label{fig:tree2}
\end{figure}

The results for the trustee display a rather different pattern.  The various regression methods, including the interdependence variables, generally performed much better than the baselines \cite{Ert2011}, especially on the estimation set, but only carefully optimized equality-reciprocity (ERC), SVM, and GPR algorithms could outperform the subgame equilibrium during prediction. All methods are dominated by the subgame perfect equilibrium but the linear regression analysis (Fig. \ref{fig:lin2}) and tree-based methods (Fig. \ref{fig:tree2})) revealed that the next most important variable is mm2, when the trustee maximizes the payoffs of the weakest player. The tree-based methods showed that this is especially important when they are already inclined to fulfill trust (b1$>0.75$).  This may also explain why equality-reciprocity or Rabin's kindness algorithms prove so strong on the estimation set– equality/fairness is incorporated into the trustworthiness decision but only after narrow self-interest is considered. While the interdependence parameters are not invoked in these models, the regression tree (not shown) does suggest that the trustor’s control over the trustee (FC$_B$) may play some small role in encouraging cooperation when the subgame perfect equilibrium tends toward defection.

\subsection{Discussion}
Based on these results it appears that the interdependence-based models best capture the response of the trustor compared to all other strategies and methods from a purely modeling perspective. This result further validates the theoretical development of the trust index in field experiments \cite{gottman2011science}. It also lends credence to our commitment model \cite{razin2019}; initially derived from field experiments \cite{rusbult_93,gottman2011science}, the commitment model is replicating well in the lab.  We also see some strong support for common-sense theories that often get less play in the game theory or HRI trust literature. This is especially true of the unilateral control each agent has in incentivizing or penalizing the other (FC), not just as a strategy but as a second-order consideration for both players \cite{Battigalli2005}. 

These second-order beliefs, that is the beliefs the trustor has of what the trustee believes of them, and vice versa, have been considered by game theory for many years, but understandably have garnered less attention from HRI. Yet it is precisely these `auxiliary' beliefs that trustors must consider. The trustee reveals them in their extraneous use of fairness, helping the weaker player/reciprocating equality when it is already in their best interest to fulfill trust. Furthermore, the trustor seems to place the commitment of the trustee before any other considerations (RC$_B$ being the root of their regression tree and the highest weighted term in the linear regression). This is only furthered by the trustee's beliefs concerning the power of the trustee in sanctioning or incentivizing the trustor.

If games had been excluded based on the Temptation condition, then the role of the trust index and the trustee's kindness would have been completely obscured, as then RC$_B<0$ and b1=0. This scenario would likely be extremely rare in human-machine trust but its further implications are left to future work.

One weakness of this data set, and game-theoretic trust in general, as well as much HRI work on trust, is the assumption that the trustor's and trustee's payoffs are equal if the trustor chooses not to trust in the first place (A$_{21}$=A$_{22}$ and B$_{21}$=B$_{22}$).  From a technical perspective, this prevents us from determining whether it is RC$_B$ or BC$_B$ that matter for the trustor and confounds whether BC$_A$ does have a significant effect due to its somewhat synthetic `perfect correlation' with FC$_A$, and thus with TI.  As discussed in \cite{razin2019}, in psychological game theory and interdependence theory, where psychological costs such as regret and satisfaction are included, these values are rarely equal.  New potential relationships generally mean regret is more costly (lowering A$_{21}$), whereas one-off interactions with strangers and team projects generally value satisfaction with coordination more highly (raising A$_{22}$). Both of these situations lead to BC$_A>$FC$_A$ \cite{razin2019}. In comparison, one's overall optimism or brand trust can increase A$_{21}$, and conversely, pessimism can decrease A$_{22}$, such that A$_{21}>$A$_{22}$. This conclusion is reasonable, as commitment is relative to the specific relationship and would thus be moderated by one's overall sense of other potential trust relationships \cite{razin2019}.  

An illustrative example further validating these points can be found in Dunning et al. \cite{dunning2014trust}.  That series of experiments looked at trustor's risk tolerance, whether they wanted to trust, felt like they should trust, and the guilt and agitation they anticipated feeling at not trusting (when the other may be trustworthy). As in many other studies, they found that people ``over-trusted" based on rationality (subgame perfect equilibria) and risk tolerance alone. In general, the choice to trust was closer to what people felt like they should do vs. what they wanted to do. This choice was therefore understood to be partially motivated by anticipated agitation at not trusting, as well as perceived approval of normative behavior from authority. From our results, we can understand both of these unilaterally as increasing the trustor's commitment. Familiarity increased repayment expectations, seemingly through the improved calibration of the threshold for FC$_A$. Furthermore, they found that when the trustee is seen as making a thoughtful decision to trust instead of just choosing at random, they are more likely to be trusted, illustrating the importance of second-order considerations. However, participants also preferred to give others the opportunity to be trustworthy, which they perceived as a sign of respect for autonomy.  In our experiment, this may point to the small effect of equality/fairness in amplifying the trustee`s SPE.  Once trustworthiness is called for, it pays to more strongly signal one`s commitment to fairness/equality as the trustee.  Further,  evidence from Dunning et al.`s trials pointed to trusting above rationality to be predicated on self-perceived moral norms of fulfilling one's social duty and to avoid casting aspersions on another's character. However, taken together these last points posit an alternative account that would suggest a key testable difference in modeling human-human vs human-robot trust interactions as norm fulfillment.

\section{Experiment 2: Breaking Down Trust}
This experiment looked to rectify the shortcomings of the previously tested data set by (a) considering trust between humans as well as between humans and various technologies, (b) employing more realistic scenarios, (c) taking into account various types and quantities of risk, and (d) breaking the $FC_A=BC_A$ and $RC_B=BC_B$ assumption of previous game theory and HRI trust research.

\subsection{Experimental Procedure 2}
In this experiment, 34 different scenarios were composed across 9 different types of risk: physical, psychological, social, time loss, performance, financial, ethical, privacy, and security, based on \cite{Stuck2020}. We assigned each participant 8 scenarios, drawn from 2 of the 9 risk types, with an equal balance of human-machine and human-human scenarios, leading to a 2x2x2 within- and between-subjects design. Examples of human-human trust included taking a friend's suggested route to avoid traffic, having a stranger watch luggage briefly, dividing up work with classmates, and participating in pharmaceutical trials. Examples of human-machine trust included following GPS guidance, using a dating app, driving an autonomous vehicle, taking emergency guidance from a robot during a fire, and trusting enemy classification from a military drone. 

Each scenario was composed of two elements: a payoff table and a scenario written out in prose.  Payoffs were created randomly but some scenarios dictated certain constraints, beyond those of Eq.\ref{eq:3}, that we coded for. The general nature of these constraints will be discussed below. Scenarios were also designed to reflect a wide range of scales ($10^0-10^7$).  To maintain consistency, participants only acted as trustors.  Given the high level of convergence in human trustee behavior in Experiment 1 to the SPE, trustee behavior for both human and machine scenarios was algorithmically determined with some noise injected. 


Sixty participants took part in this experiment (55\% male, 43\% female, 2\% non-identified), ranging from 18-50+ (85\% between 18-39), and 78\% having at least some post-secondary education.  Before the experiment, participants underwent training including a practice round to become familiarized with the layout, expectations, and most importantly, how to read and understand the payoff table.  Their understanding of gameplay was assessed both after training and at the end of the experiment. After the experiment, general feedback was solicited and a number of insights into carrying out such experiments in the future were collected. Given our desire that participants understand each task, there was no time limit, and most spent 2-4 minutes per scenario.

The experiment was carried out using the Gorilla Experiment Builder (\url{www.gorilla.sc}) to create and host our experiment \cite{anwyl2020gorilla}. All research performed with human participants was done in compliance with all relevant national regulations, institutional policies and in accordance with the tenets of the Helsinki Declaration, and was approved by the Georgia Institute of Technology's IRB. Participants were recruited through Prolific, and the data was collected between March 22-March 23, 2021.

All the same algorithms deployed in Experiment 1 were tested again here, with the exception of Gaussian Process Regression which was replaced with binomial regression. All algorithms were modified to accommodate $A_{21}\neq A_{22}$ and $B_{21} \neq B_{22}$. Furthermore, whereas earlier we had a regression problem to solve, now with every participant having different payoffs, we approached the experiment as a classification problem.  While this will affect the meaning of the error rate, the overall patterns of performance and variable importance should remain clear.

\subsection{Results}
In this experiment, we only modeled the trustor and not the trustee. Thus, we did not have to consider the final three of the ``Seven Strategies" from Table \ref{tab:vars}.  VIF was once again performed, resulting in maxmin, lev1, and $RC_A$ being dropped, keeping the remaining VIFs $<3.5$). Inequality Aversion, Equality Reciprocity, Charness-Rabin, and the SPE were all strongly correlated with each other (corr=0.65-0.83) and exhibit multicollinearity, so only the SPE was retained.

\begin{table}
    \centering
    \begin{tabular}{Sl|c|c|c|l}
     &  \multicolumn{3}{c}{Estimation}\\
      Method  & Total & H-H & H-M  \\\hline
      \parbox{3cm}{Subgame Perfect\\ Equilibrium (SPE)}  & 0.446 & 0.396&0.489\\\hline
      Inequality Aversion &0.435 &0.454 &0.421\\\hline
      \parbox{3cm}{Equality Reciprocity\\ (ERC)} & 0.435 & 0.394&0.470 \\\hline
      Charness-Rabin (CR) & 0.442& 0.394& 0.481 \\\hline
      Seven Strategies & 0.410 & 0.361&0.421 \\\hline 
      Linear Reg. & 0.348 & 0.306&0.333\\\hline
      Binomial Reg. & 0.348&0.310&0.307\\\hline
      Clas. SVM  & \cellcolor{ltblue}\textbf{0.079} & \cellcolor{ltblue}\textbf{0.093}&\cellcolor{ltblue}\textbf{0.095}\\\hline
      Clas. Tree & \cellcolor{ltblue}\textbf{0.117}&\cellcolor{ltblue}\textbf{0.134}&\cellcolor{ltblue}\textbf{0.102}\\\hline
      Clas. KNN Ensemble& \cellcolor{ltblue}\textbf{0}&\cellcolor{ltblue}\textbf{0}&\cellcolor{ltblue}\textbf{0}\\\hline
          \end{tabular}
       \caption{Mean squared error for trust and trust fulfillment: All classifications were run at least initially with all Seven Strategies and interdependence variables. H-H are human-human trust scenarios and H-M are human-machine. The three best performers along each category are highlighted.}
    \label{tab:main_results2}
\end{table}

\begin{table}
    \centering
    \begin{tabular}{l|c|c|c|l}
     Method  & ROC-AUC & MCC & k-Fold Loss\\\hline
     SPE &   0.51    & 0.085 & 0.446\\\hline
     IA &    0.53 & 0.121     & 0.492 \\\hline
     ERC &   0.53 & 0.111   & 0.442  \\\hline
     CR &    0.60  & 0.093 & 0.417\\\hline
       Binomial Reg.$^*$ &  \cellcolor{LimeGreen!60}\textbf{0.70}& \cellcolor{Lavender!50}\textbf{0.356} & \cellcolor{ltblue}\textbf{0.348} \\\hline
        Clas. SVM$^*$  & \cellcolor{LimeGreen!60}\textbf{0.66} & \cellcolor{Lavender!50}\textbf{0.258} & \cellcolor{ltblue}\textbf{0.387}\\\hline
       Clas. Tree  & 0.58  & 0.104 & 0.419\\\hline
       Clas. Tree Ensemble$^*$  & \cellcolor{LimeGreen!60}\textbf{0.70} & \cellcolor{Lavender!50}\textbf{0.293} & 0.395\\\hline
       Clas. KNN Ensemble  & 0.62  & 0.205 & \cellcolor{ltblue}\textbf{0.379}\\\hline
              \end{tabular}
     \caption{Performance measures: the receiver operating characteristic's area-under-the-curve (AUC), the Matthew's correlation coefficient (MCC), and the k-fold loss. The * indicates these classifiers were optimized to minimize k-fold loss using the estimator as a baseline. The three best performers along each category are highlighted.}
        \label{tab:main_results4}
\end{table}

The total variance that could be explained by the SPE alone was 55.6\%. Like in Experiment 1, the strategies from game theory only performed slightly better than the SPE, at best.  The linear and binomial regressions worked somewhat better than these strategies, reaching 75\% accuracy.  However, the machine learning classifiers performed estimation significantly better, all achieving over 85\%. These classifiers were tested on the Interdependence terms and indices as well as a set combining the Seven Strategies with the Interdependence terms. Once again, the classifiers all performed as well or better with the Interdependence terms alone, with the SVM reaching a maximum of 92\% accuracy and the ensemble KNN 100\% for estimating the decision to trust based on the Interdependence terms alone over the whole data set.  The estimations for human-human vs human-machine were extremely similar across the board, with the more traditional game-theoretic strategies and regressions performing somewhat better for human-human. This pattern did not replicate for the classifiers using interdependence-only terms.  All of these results are summarized in Table \ref{tab:main_results2}.

\begin{figure*}[t]
    \centering
    \includegraphics[width=2\columnwidth, trim = 1cm 2cm 1cm 4cm,clip]{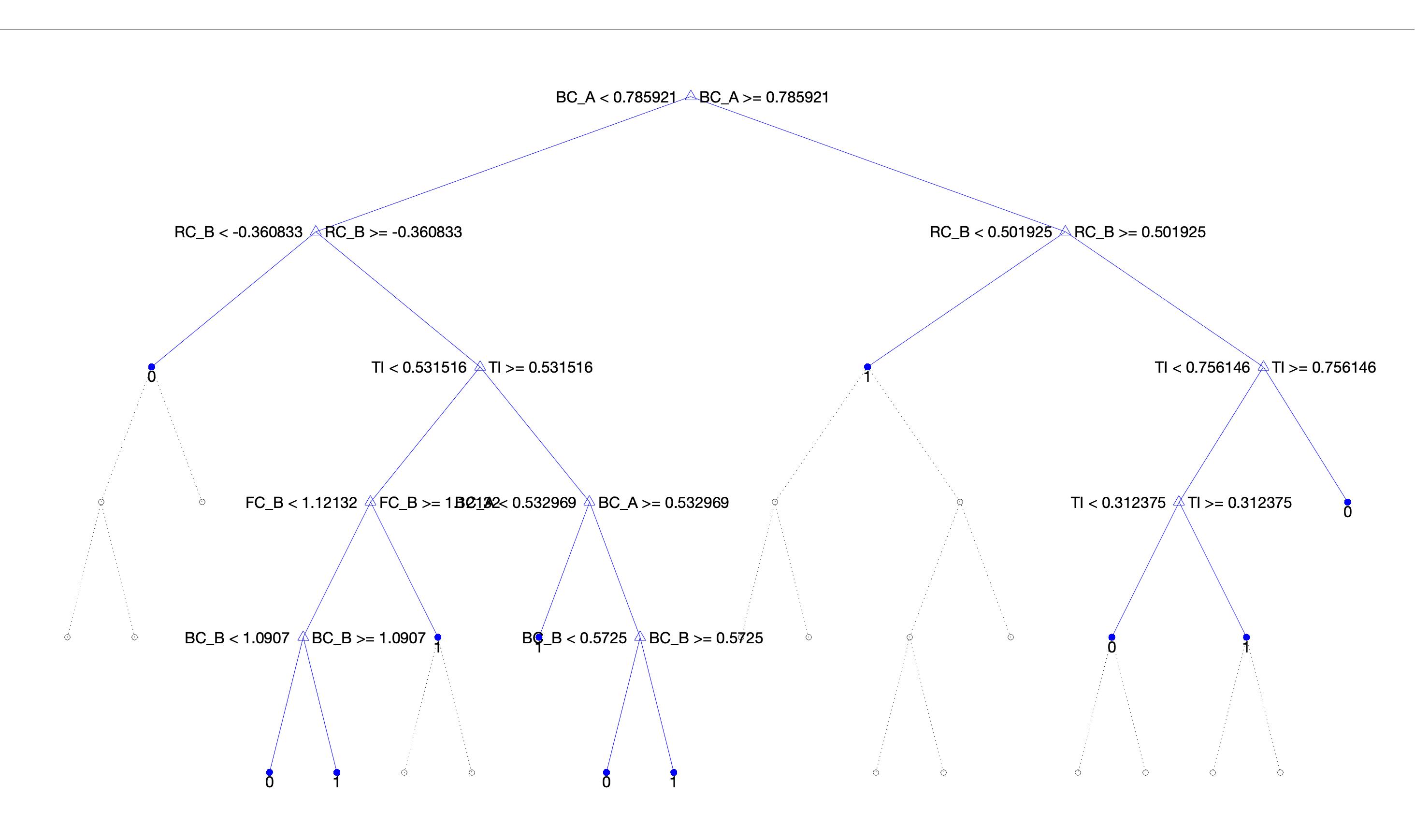}
    \caption{Full classification tree for when the trustor decides to trust from Experiment 2. Pruning was performed during optimization. Leaf nodes marked with 0`s indicate lack of trust and those with 1`s indicate trust.}
    \label{fig:tree5}
\end{figure*}

\begin{figure}[h]
    \centering
    \includegraphics[width=\columnwidth,trim=7pt 6pt 45pt 10pt,clip]{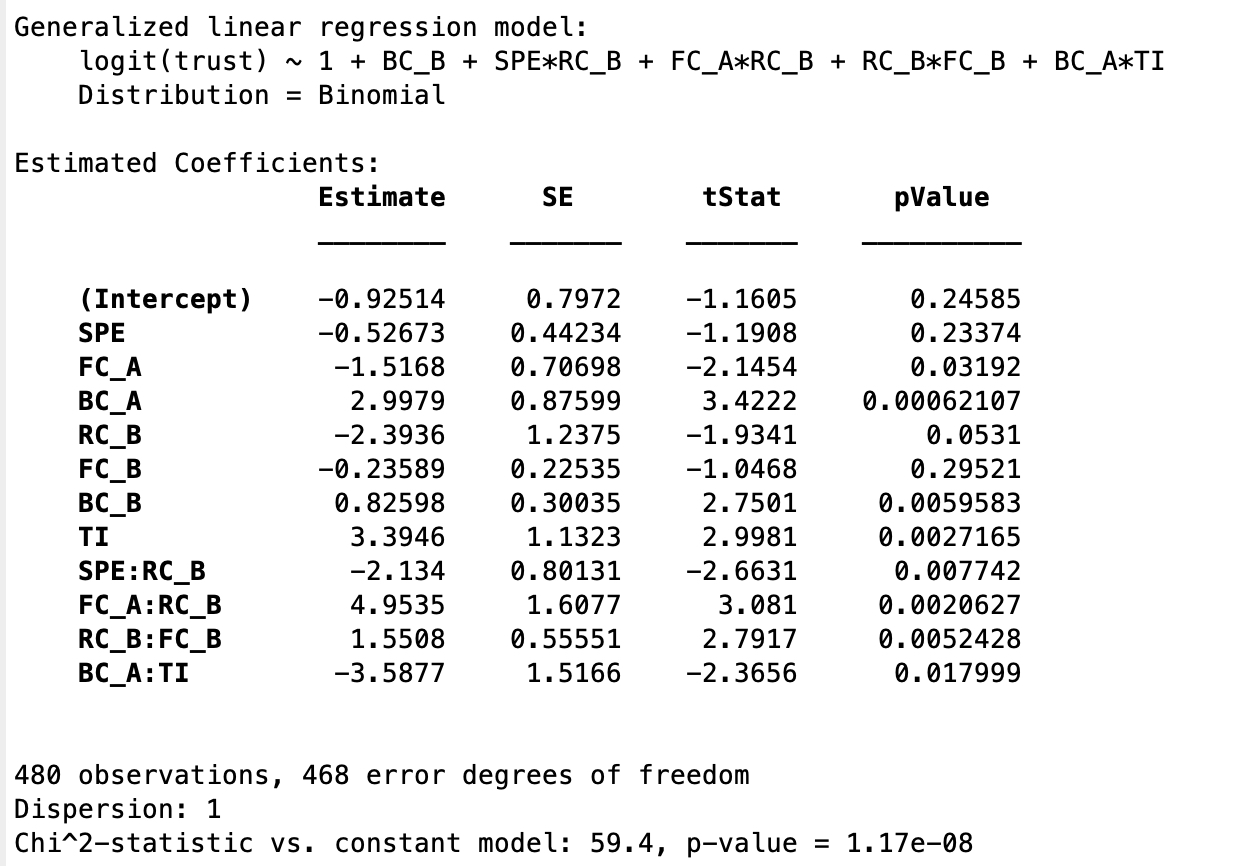}
     \caption{Step-optimized binomial regression. Optimization led to all non-interdependence variables to be dropped except the SPE.}
     \label{fig:reg2}
\end{figure}

We calculated other performance measures, such as the receiver operating characteristic area-under-the-curve (ROC-AOC), Mathew's correlation coefficient (MCC), and k-fold loss, to assess model fit and ability to predict trust (see Table \ref{tab:main_results4}). They indicate the same general pattern of interdependence dominating the more traditional measures, however, it is clear, though not surprising, that prediction errors are significantly higher than estimation errors. Most of the game-theoretic approaches did not do much better than random chance at prediction (estimated through k-fold loss), whereas the interdependence approaches all improved upon that.  Surprisingly, binomial regression edged out the Classification Tree as the third-best performer. The binomial regression (Fig. \ref{fig:reg2}) and SVM were further optimized to minimize k-fold loss and the classification tree was improved upon by leveraging an ensemble learner.

\subsection{Discussion}
\subsubsection{When to Trust: Part II}
Once again the interdependence variables were retained and dominated across the board throughout the machine learning methods, especially the KNN ensemble and SVM.  As in Experiment 1, the SVM did not allow us to `see inside' and understand which variables mattered most. However, we can turn to the classification tree to understand not only the significant variables but even the underlying logic (Fig. \ref{fig:tree5}).  Except for the root, this tree resembles that of Experiment 1, with a slightly negative RC$_B$ ranking above TI. While the split values differ somewhat they approximate the pattern we saw before.  The differences here are that we have now disentangled FC$_A$ and BC$_A$, as well as RC$_B$ and BC$_B$.  While before BC$_A$ was subsumed into TI through its forced equivalence to RC$_A$, now we can see that BC$_A$ is important enough to become the root of this tree. BC$_B$ and FC$_B$  also serve as interesting additions as they were seemingly absent from the tree in Experiment 1 despite being indicated as playing a role in the linear regression there.

The power of the classification tree is the explainability it enables. Working through it, we see that the trustor's primary consideration is BC$_A$ or what they stand to gain from cooperation, followed by the trustee's commitment (RC$_B$), followed by the Trust Index (TI). This alone gets us to 70\% accuracy, well above the other methods.  Further refinements also account for FC$_B$ and BC$_B$, while the whole tree taking on a finer and finer grain. FC$_A$ by itself is by far is the least important variable of the interdependence terms for determining when the trustor trusts.  However, it has a pivotal role in TI and it's clear that the main roles of FC$_A$ (and even RC$_A$) is through its interaction in TI.

\subsubsection{The Limits of Realism}
In this experiment, subjects were presented with a text-based scenario as well as a payoff table and asked if they would choose to trust or not, whereas when the first data set was built and tested only payoff tables were provided without much grounding in reality. However,  a number of valid concerns arose in our more realistic approach to trust problems. The amount of time and focus people placed on the table vs. the text was a serious concern as was the legibility of the table. The first experiment used participants in a university game theory class who were used to reading payoff tables, whereas the participants in the second experiment were drawn from a general pool, over a wide age range (18-50+) and with varying degrees of education.  After the experiment, participants were asked to assess their experience.  All said that the amount of training was sufficient but some felt it was overly wordy. Others indicated that the table could be confusing, as it simply presented so much information, and that a redesign would help. We expected this issue and thought it would lead many to focus on the text-based scenarios. However, the majority of respondents said that they focused more on the table, as it provided and summarized key information, especially the more scenarios they went through.  This indicates that a longer training period may be appropriate. 

The need for a longer training period, the split focus brought by increasing realism, and the shift from a regression problem to a classification problem all increased noise in the data. This, in turn, may help shed light on why the accuracy of all approaches fell significantly between Experiment 1 and Experiment 2. Even so, the overall pattern of the findings remained the same.

Given the level of understanding we aimed to achieve and the novelty of the scenario format, no time limit was instantiated, as mentioned above.  Thus, while many types of risk were tested, this experiment did not test trust under time pressure. Time pressure acts not only as another category of risk but one that directly impacts cognitive workload.  This is left for future research, though whether it can be realistically tested within this framework is an open challenge.

Another key takeaway from increasing realism revealed itself when we were designing the scenarios.  Some scenarios indicated implicit payoff constraints that we had to account for when generating the payoff matrices.  The most common of these were $A_{22}>A_{21}$ (11/34 scenarios),  $B_{22}>B_{21}$ (10/34 scenarios), and $B_{11}>B_{12}$ (7/34 scenarios). Generally, this meant that often in trust interactions, satisfaction for not trusting the untrustworthy outweighed the regret for not trusting the trustworthy (as discussed in Sec. 3.5), that suspicion hurts potential trustees, and that the Temptation condition does not always even apply in real life (see Sec. 1.3).  

A final concern regarding realism is while machine type was specified to some degree in each scenario (\textit{autonomous vehicle, emergency guidance robot}, much was left to the participants' imagination including the extent of anthropomorphism.  While this avoided anchoring bias that showing pictures or more detailed descriptions may have introduced, it leaves open questions regarding the influence of design on perceived risk and situation normality.

\section{General Discussion}
\subsection{A Declaration of Interdependence}
Together, our two studies revealed the power of the interdependent approach to understanding what defines trust games and when trust actually occurs.  Game theory played a crucial role in helping define and refine this approach.  Both interdependence theory and game theory converged on how they defined what constitutes a trust game.  The requirements of exposure and improvement are accepted across the board, and interdependence theory allowed us to understand those requirements more deeply, as a set of constraints on commitment, cooperation, and coercion.

However, once it came to how people actually play trust games, the game theory strategies proved insufficient, especially once applied to real-life scenarios. It remains clear that people depart significantly from `rational' gameplay of the SPE, a fact which both HRI and psychological game theory have long struggled to explain. Part of the problem may be in the more narrow definition of a trust game in game theory, specifically the Temptation condition ($B_{12}>B_{11}$). In fact, in the second experiment, we tested Temptation explicitly against our other variables and found that it has a near-perfect negative correlation (-0.96) with the SPE, clearly demonstrating that the standard game theory set up has a deep internal contradiction - requiring Temptation while hypothesizing the importance of SPE simultaneously. It is little wonder that `rationally' one should never trust.

Thus, while it is tempting to understand Temptation at what makes trust games actual `games', it is crucial that here we confirmed that it is neither a necessary nor sufficient condition as such. Once we broke the $B_{21}=B_{22}$ equality, Temptation could be re-framed as either $RC_B<0$ or $BC_B<0$, that is the relative power/lack of commitment of the trustee or their gains from being competitive. While trust is harder when $RC_B<0$ or $BC_B<0$ it is not impossible, nor do $RC_B>0$, $BC_B>0$  guarantee that trust will be given or reciprocated. The game is still afoot.

Beyond justifying the dropping of this constraint, the interdependence approach provides a better explanation of trust, both in terms of accuracy and parsimony.  Crucially, interdependence allows us to have a conversation about HRI trust without having to resort to reciprocity, altruism, and fairness backed by convoluted explanations of anthropomorphism.  By using commitment, coercion, and cooperation to explain trust our model allows us to bridge the divide between human-human and human-machine trust. Furthermore, this approach couches trust in familiar terms, those that we use regularly to describe when and why humans trust.

\subsection{Implications for Human-Robot Interaction}
\label{sec:HRI_imp}
If trusting is about perceived commitment and cooperative gains as opposed to strict rationality, fairness, equality, or respect then very different conclusions may be drawn regarding trust as it relates to humans vs. robots.  Robots are already perceived by humans as being more fair, just, and even reliable \cite{deVisser,hauslschmid2017supportingtrust}, though this effect is moderated by anthropomorphization. Thus, per Experiment 1, robots conform nicely to the notion of the rational trustee, following the SPE, even if they don't have exhibit extra marginal gains from reciprocity or fairness (mm2).  On the other hand, it then falls to the trustor (or modeler) to capture how much benefit the robot can bring the trusting human either unilaterally (FC$_A$) or through cooperation (BC$_A$), as well as how much they should commit (RC$_A$) to the interaction or consider alternatives (CL$_{alt}$).  Furthermore, it appears that the ``over-trust" of robots and humans may really come down to perceived gains, power, and need. In the motivating example of the human and the self-driving car, the relative assessments of commitment, safety (via coercion), and reputation (via cooperation) seemed to explain the interaction more effectively than kindness or fairness.

Several major dimensions of HRI trust can be understood through the lens of interdependence. First and foremost, there are close parallels between reflexive, fate, and bilateral control and the recently proposed and aptly named autonomy dimensions of Commitment, Specification, and Control of \cite{atkinson2014shared}. More specifically, interdependence can be seen as another set of insights into the antecedents, correlates, and underlying dimensions of human-robot trust.  Affective trust (often called benevolence), assessing whether the other agent is competitive or wishes to cooperate (as in \cite{Gefen2003,schaefer2013perception,Chien2014}), is foundational to determining whether a trust game even exists \cite{razin2019} and underlies bilateral control (BC). Social \cite{jian,schaefer2013perception} and structural trust \cite{McKnightD2011,Gefen2003} are key to determining levels of commitment (RC) and the norms at play (e.g. equity or kindness). Familiarity \cite{Korber2004,Gefen2003} helps establish thresholds and refines calibrations of anticipated payoffs. While anthropomorphism may shift strategy choice (especially for trust repair) by triggering psychological norms \cite{nass1999people,Hoff2015}, it is likely to also play a key role in establishing familiarity and situation normality \cite{Epley2007,Park2020,Gefen2003}, and thus feeds into trust calibration. This effect, however, may be confounded by the uncanny valley at some limit \cite{hauslschmid2017supportingtrust,Nordheim2018}. While more mechanical robots may be seen as fairer and more efficient, more humanoid ones may be accorded more respect and forgiveness during trust repair. In the middle of the ``uncanny trust valley", robots may be seen as having qualities of both ends, either for better \cite{hauslschmid2017supportingtrust} or for worse \cite{Nordheim2018}. 

However, if \cite{dunning2014trust} is correct in that humans are motivated to trust via norm fulfillment out of respect, a completely independent theory of trust would be necessary for humans vs. robot trustees. This possibility is made all the more interesting for HRI if that respect is predicated on individual moral autonomy vs a personal autonomy based on agency \cite{zhu2009intentional,alaieri2016ethical}.  On the other hand, previous work \cite{jian,lyons2012human,verberne2012trust} on trust in social psychology and HRI has suggested that the underlying dimensions, antecedents, and correlates of trust for human-human and human-robot interaction heavily overlap and function in similar ways. Our work strongly comes out on the side of the latter and leaves a major testable contention for future work. 

Finally, we have primarily focused on the human being the trustor and the robot being the trustee.  The modeling approach we have taken above, though, further opens the door to allowing robots to decide whether to trust the humans with whom they interact. Perhaps, more importantly, such models would allow robots to be more self-aware of higher-order self-reflection - being able to assess the likelihood that they will be trusted by humans and whether this trust is well-calibrated, an assessment they can use to give feedback in aiding the human to calibrate their own trust even further.

\section{Conclusion}
HRI and game theory have each been slowly working towards more complete theories, models, and metrics of trust for the last 35 years.  Both have gone beyond capability and pure rationality and started to incorporate psychological and social factors. However, these two fields have yet to fully recognize each other's potential for cross-calibration.  Interdependence theory, with its focus on cooperation, control, and commitment, is key to bridging this gap. Crucially, this work further validated previous research on interdependence theory from social psychology by testing it on a wide range of games and a large subject pool.  These variables, especially as they relate to trusting, are shown to be powerfully predictive and are equally amenable to being integrated with previous game-theoretic trust work, as well as expand on an emerging holistic approach to trust in HRI and beyond.  Interdependence-based approaches, unlike previous game theory strategies for assessing trust, are equally understandable for human and non-human agents and imply a strong general neuro-psychological model of trust, furthering our goal of illustrating a more complete theory of interactional trust for humans and automation.


\section*{Conflict of Interest}
The authors have no conflicts of interest to report.

\section*{Data Availability Statement}
The data that support the findings of this study are available from the corresponding author, YSR, upon reasonable request.



\bibliographystyle{IEEEtran}
\bibliography{biblio}
\appendix
\section{Appendices}
\subsection{Glossary of Terms}
\FloatBarrier
\begin{table}[H]
    \centering
    \begin{tabular}{lr}
      A$_{11}$ & Trustor's payoff for successfully-placed trust\\
      A$_{12}$ & Trustor's payoff (cost) if betrayed\\
      A$_{21}$ & Trustor's payoff (cost) for not trusting and regret\\
      A$_{22}$ & Trustor's payoff if they don't trust an untrustworthy player\\
      B$_{11}$ & Trustee's payoff for successfully-returned trustworthiness\\
      B$_{12}$ & Trustee's payoff if they betray the trustor\\
      B$_{21}$ & Trustee's payoff (cost) if they are not trusted\\ & when they are trustworthy\\
      B$_{22}$ & Trustee's payoff (cost) if they are not trusted\\ & when they are not trustworthy\\
      RC & \textit{Reflexive control}: How much unilateral power\\ & each actor has over their own outcomes\\
      FC & \textit{Fate/Partner control}: How much unilateral power\\ & each actor has over the other player's outcomes\\
      BC & \textit{Bilateral control}: How much one actor's choice\\ &further facilitates
       or inhibits the other's outcomes,\\ & \textit{i.e.} the cooperative gain\\
      $\tau_B$ & Probability of the trustee acting trustworthy\\
      \textit{TI} & Gottman`s Trust Index\\
      CL$_{alt}$ & Comparison level for the alternative\\
      SPE & Sub-perfect Equilibrium\\
    \end{tabular}
    \label{tab:glossary}
\end{table}

\subsection{Derivations of Interdependent Trust Conditions}

\textbf{Theorem 1} If Improvement and Exposure are true (Eqs. \ref{eq:1}, \ref{eq:3}), $BC_A>0$\\

\textit{Given $A_{11}>A_{21}$ and $A_{22}>A_{12}$}
\begin{align*}
    A_{11}>A_{21}\cap&A_{22}>A_{12}\\
    A_{11}+A_{22}>&A_{21}+A_{12}\\
    A_{11}+A_{22}-&A_{21}-A_{12}>0\\
    \therefore\: BC_A>0
\end{align*}

\textbf{Theorem 2} If Improvement and Exposure are true, $FC_A>0$\\

\textit{Given $A_{11}>A_{22}$ and $A_{21}>A_{12}$}
\begin{align*}
    A_{11}>A_{22}\cap&A_{21}>A_{12}\\
    A_{11}+A_{21}>&A_{22}+A_{12}\\
    A_{11}+A_{21}-&A_{22}-A_{12}>0\\
    \therefore\: FC_A>0
\end{align*}

\textbf{Theorem 3} If Improvement and Exposure are true, $FC_A\geq|RC_A|$\\

\textit{Given $A_{21}>A_{12}$}
\begin{align*}
A_{21}>&A_{12}\\
2A_{21}>&2A_{12}\\
A_{21}-A_{12}>&A_{12}-A_{21}\\
A_{21}-A_{12}+A_{11}-A_{22}>&A_{12}-A_{21}+A_{11}-A_{22}\\
2FC_A>&2RC_A\\
\therefore\: FC_A>&RC_A\\
\\
A_{11}>&A_{22}\\
2A_{11}>&2A_{22}\\
A_{11}-A_{22}>&-A_{11}+A_{22}\\
A_{11}-A_{22}+A_{21}-A_{12}>&-A_{11}+A_{22}+A_{21}-A_{12}\\
2FC_A>&-2RC_A\\
\therefore\: FC_A>&-RC_A\\
\\
A_{11}>&A_{22}\\
2A_{11}>&2A_{22}\\
A_{11}-A_{22}>&-A_{11}+A_{22}\\
FC_A>RC_A \cap& FC_A>-RC_A\\
\therefore\: FC_A>|RC_A|
\end{align*}

\textbf{Theorem 4} If Improvement and Exposure are true, $BC_A\geq|RC_A|$\\

\textit{Given $A_{21}>A_{12}$}
\begin{align*}
A_{22}>&A_{12}\\
2A_{22}>&2A_{12}\\
A_{22}-A_{12}>&A_{12}-A_{22}\\
A_{22}-A_{12}+A_{11}-A_{21}>&A_{12}-A_{22}+A_{11}-A_{21}\\
2BC_A>&2RC_A\\
\therefore\: BC_A>&RC_A\\
\\
A_{11}>&A_{21}\\
2A_{11}>&2A_{21}\\
A_{11}-A_{21}>&-A_{11}+A_{21}\\
A_{11}-A_{21}+A_{22}-A_{12}>&-A_{11}+A_{21}+A_{22}-A_{12}\\
2BC_A>&-2RC_A\\
\therefore\: BC_A>&-RC_A\\
\\
BC_A>RC_A \cap& BC_A>-RC_A\\
\therefore\: BC_A>|RC_A|
\end{align*}

\subsection{Derivations of Trust Measures}
\textbf{Theorem 1} The Nash equilibrium can be expressed in Interdependence terms as $\tau_B=\frac{1}{2}-\frac{RC_A}{2BC_A}$
\\
\\
\textit{Given $BC_A=0.5(A_{11}+A_{22}-A_{12}-A_{21})$ and $RC_A=0.5(A_{11}+A_{12}-A_{12}-A_{22})$}
\begin{align*}
     \tau_B &= \frac{A_{22}-A_{12}}{A_{11}+A_{22}-A_{12}-A_{21}}\\
             \\
            &= \frac{A_{22}-A_{12}}{2BC_A}\\
            \\
             &= \frac{A_{22}-A_{12}}{2BC_A}\times\frac{2}{2}\\
             \\
              &= \frac{2A_{22}-2A_{12}}{4BC_A}\\
              \\
               &= \frac{A_{11}-A_{11}+A_{21}-A_{21}+2A_{22}-2A_{12}}{4BC_A}\\
               \\
               &= \frac{(A_{11}+A_{22}-A_{12}-A_{21})-(A_{11}+A_{12}-A_{12}-A_{22})}{4BC_A}\\
               \\
               &= \frac{2BC_A-2RC_A}{4BC_A}
             \\
             \\
     \therefore\: \tau_B&=\frac{1}{2}-\frac{RC_A}{2BC_A}.
\end{align*}
\textbf{Theorem 2} Gottman's trust index can be expressed in Interdependence terms as $TI=\frac{1}{2}-\frac{RC_A}{2FC_A}$
\\
\\
\textit{Given $FC_A=0.5(A_{11}+A_{21}-A_{12}-A_{22})$ and $RC_A=0.5(A_{11}+A_{12}-A_{12}-A_{22})$}
\begin{align*}
     TI =& \frac{A_{11}-A_{22}}{A_{11}+A_{21}-A_{12}-A_{22}}\\
             \\
            =& \frac{A_{11}-A_{22}}{2FC_A}\\
            \\
             =& \frac{A_{11}-A_{22}}{2FC_A}\times\frac{2}{2}\\
             \\
              =& \frac{2A_{11}-2A_{22}}{4FC_A}\\
              \\
               =& \frac{A_{12}-A_{12}+A_{21}-A_{21}+A_{11}+A_{11}-A_{22}-A_{22}}{4FC_A}\\
               \\
               =& \frac{(A_{11}+A_{21}-A_{12}-A_{22})+(A_{11}+A_{12}-A_{12}-A_{22})}{4FC_A}\\
               \\
               =& \frac{2FC_A+2RC_A}{4FC_A}
             \\
             \\
     \therefore\: TI=&\frac{1}{2}+\frac{RC_A}{2FC_A}.\\
\end{align*}

\end{document}